\documentclass[a4paper,11pt]{article}
\usepackage{aaskaiid}
\usepackage{orcidlink}

\usepackage{xspace}
\usepackage{wrapfig}

\newcommand{\HI}{\mbox{H\,{\sc i}} }

\title{Deep \HI observations of cold gas inflow and outflow}
\ShortTitle{Deep \HI observations of in- and outflow}

\author[1,2,3]{W.J.G. de Blok\orcidlink{0000-0001-8957-4518}}
\ShortName{de Blok et al.} 
\author[4,5]{D.B. Fisher\orcidlink{0000-0003-0645-5260}}
\author[6]{K. Tachihara\orcidlink{0000-0002-1411-5410}}
\author[7,8]{F.M. Maccagni\orcidlink{0000-0002-9930-1844}}
\author[9]{J. Wang}
\author[10]{R. Enokiya\orcidlink{0000-0003-2735-3239}}
\author[11]{L. Chemin\orcidlink{0000-0002-3834-7937}}
\author[6]{T. Hayakawa}
\author[1]{D. Kleiner\orcidlink{0000-0002-7573-555X}}
\author[2]{D.J. Pisano\orcidlink{0000-0001-7996-7860}}
\author[12]{S.H. Oh}

\affiliation[1]{Netherlands Institute for Radio Astronomy (ASTRON), Oude Hoogeveensedijk 4, 7991 PD Dwingeloo, the Netherlands}
\affiliation[2]{Dept.\ of Astronomy, Univ.\ of Cape Town, Private Bag X3, Rondebosch 7701, South Africa}
\affiliation[3]{Kapteyn Astronomical Institute, University of Groningen, PO Box 800, 9700 AV Groningen, The Netherlands}
\affiliation[4]{Centre for Astrophysics and Supercomputing, Swinburne University of Technology, Hawthorn, VIC 3122, Australia}
\affiliation[5]{ARC Centre of Excellence for All Sky Astrophysics in 3 Dimensions (ASTRO3D)}
\affiliation[6]{Department of Physics, Graduate School of Science, Nagoya University, Furo-cho, Chikusa-ku, Nagoya, 464-8602, Japan}
\affiliation[7]{INAF -- Osservatorio Astronomico di Cagliari, via della Scienza 5, 09047, Selargius (CA), Italy}
\affiliation[8]{Wits Centre for Astrophysics, School of Physics, University of the Witwatersrand, 1 Jan Smuts Avenue, 2000, Johannesburg, South Africa}
\affiliation[9]{Kavli Institute for Astronomy and Astrophysics, Peking University, Beijing 100871, China}
\affiliation[10]{National Astronomical Observatory, Japan, 2-21,-1, Osawa, Mitaka, Tokyo, 181-8588, Japan}
\affiliation[11]{Universit\'e de Strasbourg, CNRS, Observatoire Astronomique de Strasbourg, UMR 7550, 67000 Strasbourg, France}
\affiliation[12]{Department of Physics and Astronomy, Sejong University, 209 Neungdong-ro, Gwangjin-gu, Seoul, Republic of Korea}

\abstract{A major question in galaxy evolution is how
  galaxies acquire sufficient gas to sustain their star formation
  rates.  \HI observations with high angular resolution and 
  sensitivity to very low column densities are some of the important
  observational ingredients that are currently still missing.  Answers
  to these questions are necessary for a correct interpretation of
  observations of galaxy evolution in the high-redshift universe and
  will provide crucial input for the sub-grid physics in
  hydrodynamical simulations of galaxy evolutions. In this chapter we
  discuss the progress that has been made over the past years, describe the various processes that lead to inflow and outflow of gas, and discuss how SKA-Mid AA4 observations can contribute to further understanding these important aspects of galaxy evolution using deep  observations of nearby individual disk and dwarf galaxies.}
  
\begin{document}
\maketitle

\newcommand{\actaa}{Acta Astron.} 
\newcommand{\araa}{ARA\&A} 
\newcommand{\aar}{A\&ARv} 
\newcommand{\aapr}{A\&ARv} 
\newcommand{\ab}{Astrobiol.} 
\newcommand{\aj}{AJ} 
\newcommand{\apj}{ApJ} 
\newcommand{\apjl}{ApJL} 
\newcommand{\apjs}{ApJSS} 
\newcommand{\ao}{Appl. Opt.} 
\newcommand{\apss}{Astro. \& Space Sci.} 
\newcommand{\aap}{A\&A} 
\newcommand{\aaps}{A\&AS.} 
\newcommand{\baas}{Bull. Am. Astron. Soc.} 
\newcommand{\caa}{Chinese A\&A} 
\newcommand{\cjaa}{Chinese J. A\&A} 
\newcommand{\cqg}{Class. Quantum Gravity} 
\newcommand{\gal}{Galaxies} 
\newcommand{\gca}{Geo. Cosmo. Acta} 
\newcommand{\icarus}{Icarus} 
\newcommand{\jcap}{JCAP} 
\newcommand{\jgr}{J. Geophys. Res.} 
\newcommand{\jgrp}{J. Geophys. Res. Planets} 
\newcommand{\jqsrt}{J. Quant. Spectrosc. Radiat. Transf.} 
\newcommand{\memsai}{Mem. SAIt} 
\newcommand{\mnras}{MNRAS} 
\newcommand{\nat}{Nature} 
\newcommand{\nastro}{Nat. Astron.} 
\newcommand{\ncomms}{Nat. Commun.} 
\newcommand{\nphys}{Nat. Phys.} 
\newcommand{\na}{New Astron.} 
\newcommand{\nar}{New Astron. Rev.} 
\newcommand{\physrep}{Phys. Rep.} 
\newcommand{\pra}{Phys. Rev. A} 
\newcommand{\prb}{Phys. Rev. B} 
\newcommand{\prc}{Phys. Rev. C} 
\newcommand{\prd}{Phys. Rev. D} 
\newcommand{\pre}{Phys. Rev. E} 
\newcommand{\prx}{Phys. Rev. X} 
\newcommand{\prl}{Phys. Rev. Let.} 
\newcommand{\psj}{Planet. Sci. J.} 
\newcommand{\planss}{Planet. Space Sci.} 
\newcommand{\pnas}{Proc. Natl Acad. Sci. USA} 
\newcommand{\procspie}{Proc. SPIE} 
\newcommand{\pasa}{PASA} 
\newcommand{\pasj}{PASJ} 
\newcommand{\pasp}{PASP} 
\newcommand{\rmxaa}{RMXAA} 
\newcommand{\sci}{Science} 
\newcommand{\sciadv}{Sci. Adv.} 
\newcommand{\solphys}{Sol. Phys.} 
\newcommand{\sovast}{Soviet Ast.} 
\newcommand{\ssr}{Space Sci. Rev.} 
\newcommand{\uni}{Universe} 

\setlength{\bibsep}{0.0pt} 

\newcommand{\kms}{km~s$^{-1}$\xspace}

\section{Ultra-deep \HI observations with SKA-Mid}

Galaxy evolution is driven by the flow of gas into galaxies, the
transformation of gas into stars, and the expulsion of gas due to the
subsequent stellar evolution processes.  Atomic neutral hydrogen (H\,{\sc i})
is an excellent tracer --- and often the main constituent --- of this
gas component, and can be observed in the 21-cm line.  The SKA will be
able to trace the gradual transformation from primordial
hydrogen into galaxies over cosmic time. 
It will provide direct and detailed observations of the
physical processes that cause this transformation and which will help to correctly
interpret this evolution.

An important process in this is gas accretion. It delivers gas from outside galaxies (either primordial or previously ejected) into the star-forming disks, which ensures galaxies can keep forming stars over a Hubble time.
In general, local galaxies only have enough gas to
sustain their SFR for few Gyr, and they must thus acquire gas from
somewhere else (see \citealt{sancisi} for an
overview). 
One of the ways in which gas can flow onto the star forming disk is through ``cold accretion'' where numerical simulations have predicted that 
gas flows in from the intergalactic medium (IGM) (or
``cosmic web'') (e.g.,
\citealt{keres}). ``Cold'' in this context means that the gas has not
been shock-heated as it entered the galaxy halo.   So far there is little
direct observational evidence for significant cold accretion. The
observed cold gas accretion in galaxies seems to be an order of
magnitude too low to explain the current star formation rates (SFR) in
galaxies \citep{sancisi,putman}.  An alternative accretion path is through the galactic fountains. These propel gas into the halo, which rains back on the disk, dragging additional gas back in \citep{Fraternali2017}. 

If cold accretion as predicted by the simulations is the dominant
process by which galaxies acquire their gas, then current
observational limits indicate it must happen at \HI column densities
below $\sim 10^{18}$ cm$^{-2}$ \citep{marasco25}, based on recent deep MeerKAT 
observations obtained as part of the MHONGOOSE (MeerKAT \HI Observations of Nearby Galactic Objects: Observing 
Southern Emitters) survey \citep{deblok24}. At these 
low column densities, the \HI\ is of course no longer the main mass component of the 
accretion flow, but serves as a tracer. 

As part of the galactic fountain process \citep{shapiro, bregman, norman}, massive stars,
through supernova explosions and stellar winds, can push gas out of
the disk and into the halo of a galaxy. This creates the holes and
bubbles frequently observed in the gas disks of galaxies
\citep{bagetakos}.  The expelled gas will cool and eventually rain
back on the disk, most likely in the form of \HI\ clouds
\citep{putman}.  Such clouds have also been observed in a number of
other galaxies as part of an extra-planar gas component
\citep{sancisi}, and presumably form the equivalent of the high and
intermediate velocity clouds (HVCs and IVCs, respectively) in our
Galaxy.  It is thought that the process of these clouds moving through
the hot gaseous halo of a galaxy provides an alternative mechanism for
accretion of gas.  Here, hot halo gas cools in a cloud's wake and is dragged
along as the cloud moves back into the disk \citep{ff13}.

The current state of the art in attempting to detect low-column density \HI\ around nearby galaxies is the MHONGOOSE survey. 
This is a MeerKAT survey of 30 nearby disk and dwarf galaxies, detecting \HI column densities down to $\sim 5 
\times 10^{17}$ cm$^{-2}$ using 55 hours of integration time per galaxies. To date this is the deepest targeted \HI\ survey of nearby galaxies. 
Some of the early results of this survey include the discovery of a large number of low-mass dwarf galaxies around our target 
galaxies. With \HI\ masses of a few times $10^6\ M_{\odot}$, these are the equivalents of Local Group dwarfs, which can now be 
detected out to $\sim 20$ Mpc with MHONGOOSE (Maccagni et al. 2024). Another early discovery is the presence of a low-column 
density outer disk component around NGC 5068. There is no clear origin for this gas, so this may indicate accretion from the 
IGM, as also hinted at by the metallicities of star forming regions (Healy et al 2024). Stacking of the MHONGOOSE data show no
strong evidence for the presence of extensive low-column density material (Veronese et al 2025). This supports the analysis by 
Marasco et al. (2025) who conclude that the infalling cool gas filaments predicted as part of the cold accretion process should 
have been detected by MHONGOOSE. This incomplete list shows the impact of the dramatic increase in quality of the MeerKAT data 
over previous studies. 

\begin{figure}[h]
    \centering
	\includegraphics[width=0.99\columnwidth]{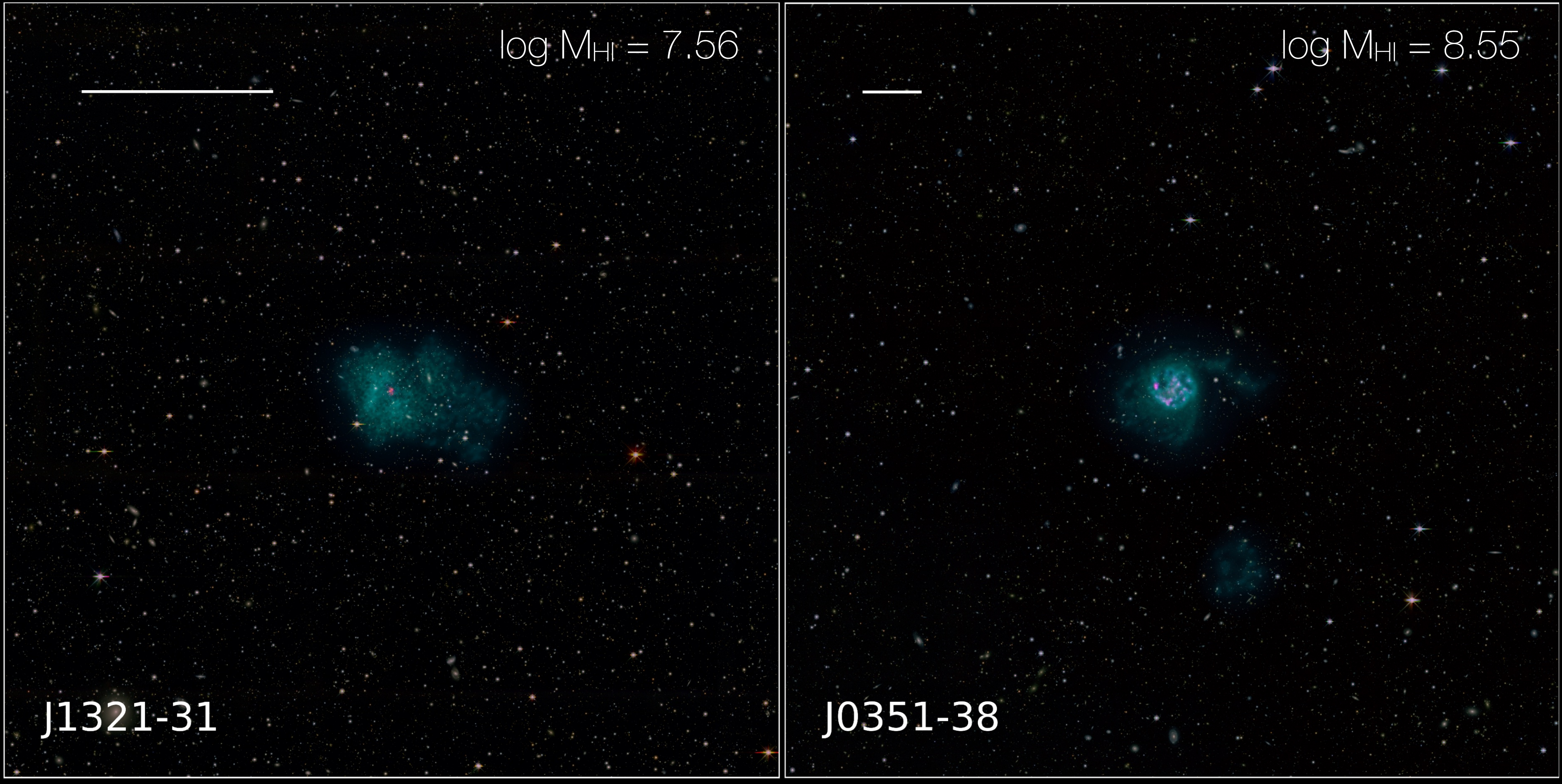}\\
    \includegraphics[width=0.99\columnwidth]{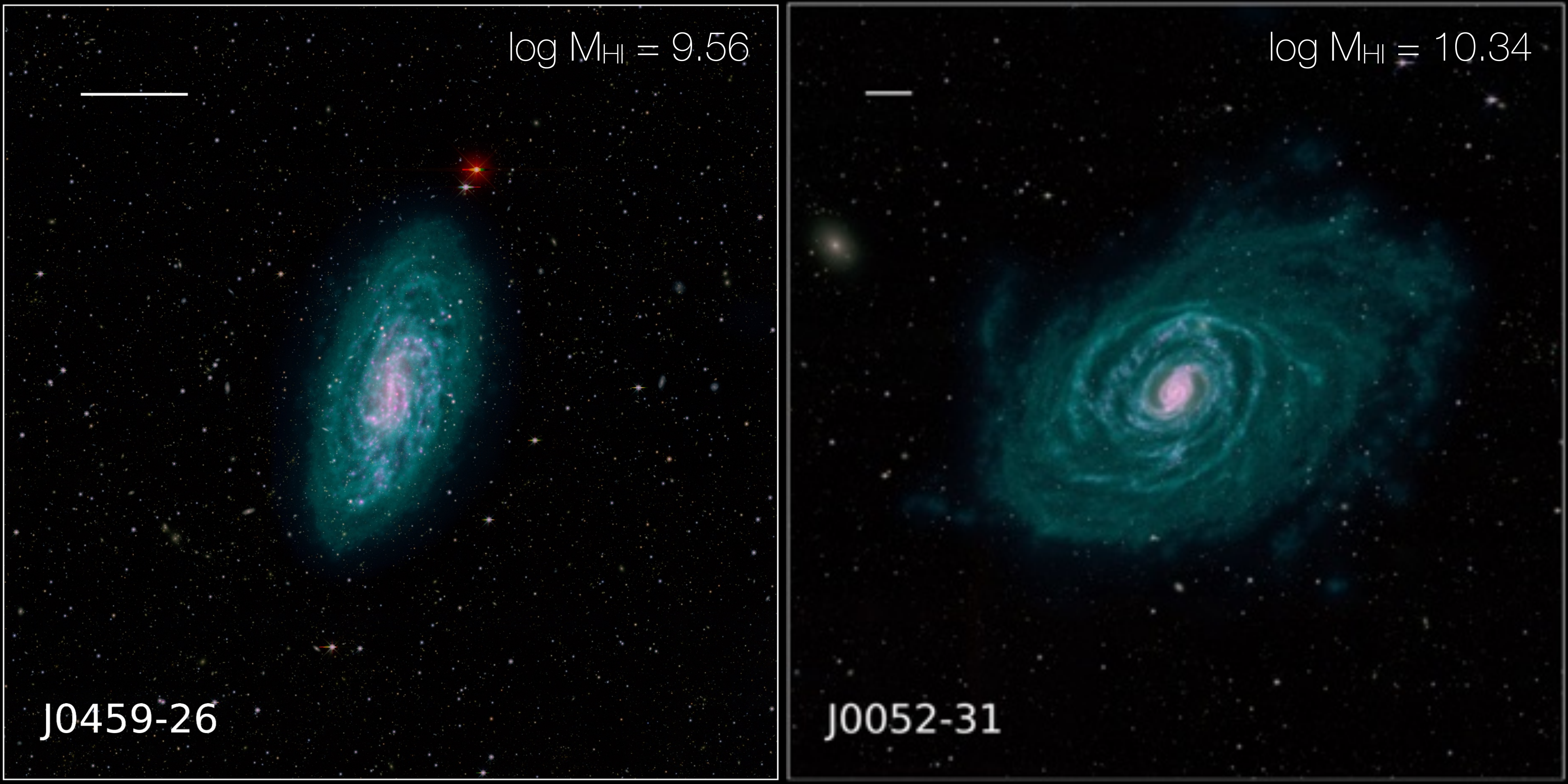}
    \caption{MHONGOOSE \HI\ distributions (cyan), combined with Galex UV (pink) and Legacy Survey optical images (background) of four sample galaxies each differing about an order of magnitude in \HI\ mass. A 10 kpc scale bar is shown in the top-left of each panel, the \HI\ mass in the top-right. The HIPASS identification is shown in the bottom-left. These correspond to (top-left to bottom-right) KK98-195, ESO 302-G014, NGC 1744 and NGC 289.}
    \label{fig:pictures}
\end{figure}

As an illustration of the quality we show in Fig.\ \ref{fig:pictures} an overview of the \HI distribution of four MHONGOOSE galaxies each 
differing by about an order of magnitude in \HI\ mass. The progression from low to high star formation rates, and the increasing 
orderedness of the disks is clearly visible. As will be discussed below, these data already give us a very clear idea of what a 
typical SKA-Mid observation of a nearby galaxy will look like.

\subsection{SKA Prospects}

\begin{figure}[h]
    \centering
	\includegraphics[width=\columnwidth]{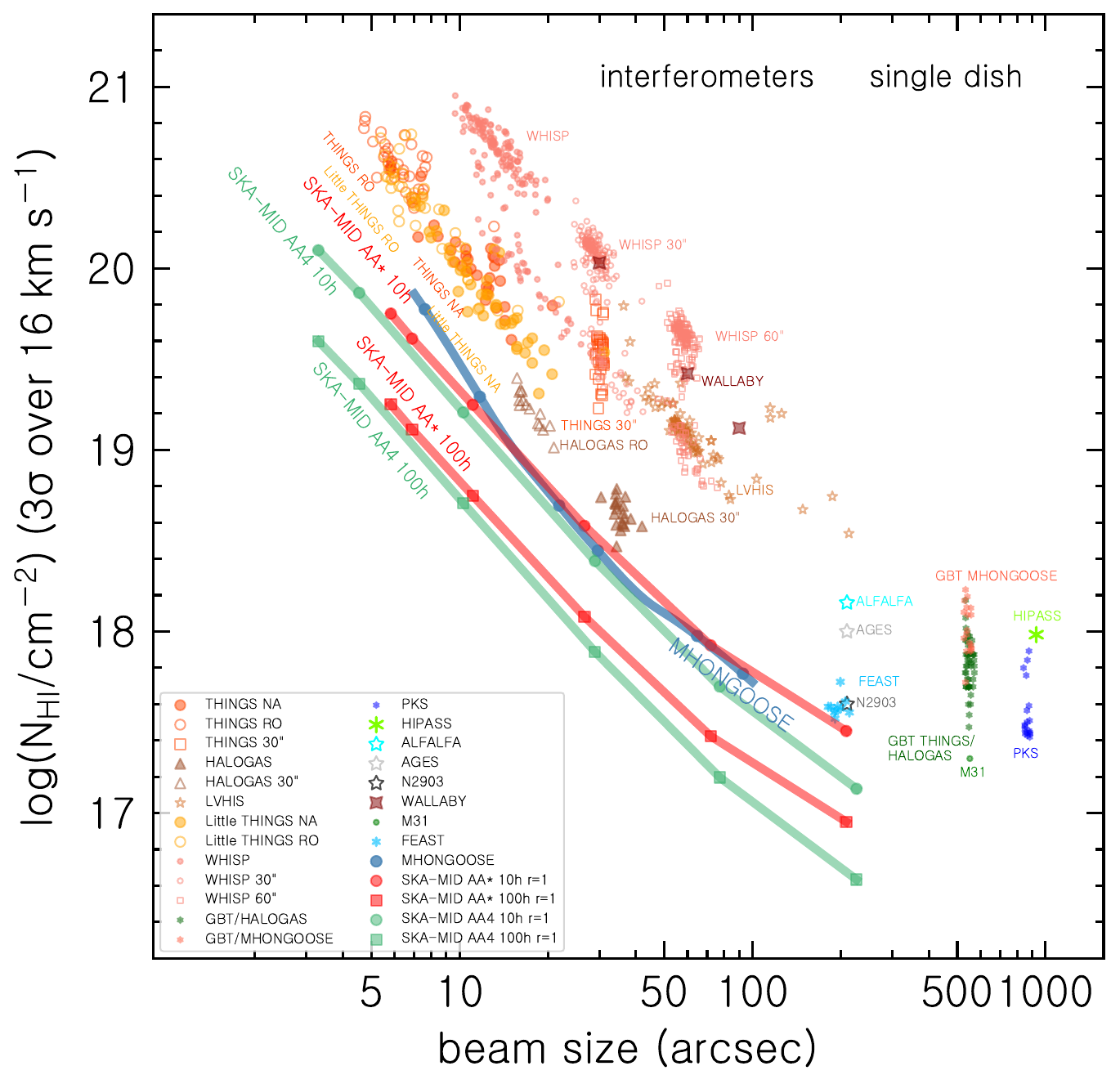}
    \caption{Sensitivity versus resolution in \HI surveys. Colored
    symbols show the $3\sigma$ column density sensitivity over 16 \kms
    for various interferometric and single-dish surveys, as indicated
    by the labels and legend and listed in the text. The thick blue
    line shows the observed MHONGOOSE sensitivities. MHONGOOSE reaches
    single-dish sensitivities but at a 10--50 times better angular
    resolution. To give an indication of the physical scales: at 10.3
    Mpc (the median distance of the MHONGOOSE sample), $10''$
    corresponds to 0.5 kpc. Galaxies that are part of a sample with a
    fixed angular resolution (THINGS $30''$, GBT, PKS, FEASTS) were given
    small, random horizontal offsets for clarity. References are given
    in the text.}
    \label{fig:surveys}
\end{figure}

In order to plan and eventually execute SKA observations it is important to have 
a good overview of the column densities and resolutions that can be reached with 
SKA in a reasonable time. 
A first indication of this was already given in \citet{deblok24} and additionally in \citet{maccagni24}. There, a comparison was 
made between (homogenized) column density sensitivities and resolutions of past 
targeted \HI surveys and those expected to be reached by SKA. In 
those papers, a preliminary SKA design was used. Here we revisit this comparison, but 
this time using the actual predictions of the SKA-Mid sensitivity calculator made 
available by SKAO. 
This comparison is shown in Fig.\ \ref{fig:surveys}.
Interferometric targeted surveys shown here are The \HI Nearby Galaxy Survey (THINGS; \citealt{walter08}), the
Hydrogen Accretion in LOcal GAlaxieS survey (HALOGAS \citealt{heald11}), the Westerbork \HI Survey Project (WHISP;
\citealt{whisp}), the Local Irregulars That Trace
Luminosity Extremes, The \HI Nearby Galaxy Survey (LITTLE THINGS;
\citealt{hunter}) and the Local Volume \HI Survey (LVHIS;
\citealt{Koribalski.2018}). Also shown is the untargeted
Widefield ASKAP L-band Legacy All-sky Blind surveY (WALLABY;
\citealt{Koribalski.2020}). For the single-dish observations we show
the column density sensitivities for the \HI Parkes All Sky Survey
(HIPASS; \citealt{Barnes.2001, Meyer.2004}), the Arecibo Legacy Fast
ALFA (ALFALFA; \citealt{Haynes.2018}), the Arecibo Galaxy Environment
Survey (AGES; \citealt{Auld.2006}), deep observations of M31
\citep{Wolfe.2016} and NGC 2903 \citep{Irwin.2009}, as well as a
number of deep GBT and Parkes observations of nearby galaxies
(\citealt{Sorgho.2019}, \citealt{Sardone.2021}, \citealt{Pingel.2018}). We also show 
deep \HI observations taken with the Five-hundred-meter Aperture
Spherical Telescope (FAST) as part of the FAST Extended Atlas of Selected Targets Survey 
(FEASTS;  \citealt{feasts}).

To ensure a proper comparison we have taken the noise per channel and
the channel widths from the source papers or the corresponding
publicly available data and homogenized these quantities to a common
channel width of 16 \kms, assuming square-root scaling of noise with
channel-width.  A 16 \kms channel width corresponds approximately to
the FWHM of an \HI\ line with a velocity dispersion of 7 \kms, which is
comparable to the lowest values seen in previous \HI\ observations of
nearby galaxies \citep{Ianjamasimanana.2017}.
Also shown are the sensitivities reached by MHONGOOSE, the deepest targeted 
\HI survey of nearby galaxies to date with 55 hours of MeerKAT observing 
time per galaxy. 
We also show the sensitivities that will be reached by the SKA-Mid AA* and 
AA4 arrays for observing times of 10 hours and 100 hours. 
The noise levels and beam sizes were derived using the early 2025 version of 
the official SKAO SKA-Mid sensitivity
calculator\footnote{{\tt https://sensitivity-calculator.skao.int/}} where we assumed a source at zero redshift (frequency 1420 MHz),  a 
spectral resolution of 1.4 \kms, with the source at a declination of $-30^{\circ}$. 
For reasons of clarity, we only show the column density sensitivities 
that will be achieved using robust weighting with robustness parameter of 1.0. 
The different resolutions shown in the Figure are achieved by tapering. Values derived using different weightings generally lie within 0.2 dex of the curves shown.
We note one difference between the SKA resolutions shown in this Figure and the ones 
derived by the sensitivity calculator. We found that for higher robust values ($
\gtrsim 0.5)$ and small beam sizes ($\lesssim 6''$) the output beam sizes as derived 
by  the calculator underestimated the size of the PSF. We re-derived the PSF based on 
the $uv$ distribution of the observations shown, and found PSF sizes that were 
typically larger by up to a factor of 2. 

There are three aspects of Fig.\ \ref{fig:surveys} that are worth noting. 
The first is the almost complete overlap of the MHONGOOSE curve and the SKA 
10h curves. Essentially, this shows that a 50 hour observation with MeerKAT is 
as sensitive as a 10 hour observation with SKA-Mid. In other words, the current MHONGOOSE observations already give us a preview of what can be expected from SKA-Mid in 10 hours in terms of \HI emission observations of the nearby universe.
The second aspect is the location of the AA* and AA4 curves with respect to each 
other. AA4 contains more dishes than AA*, but as these are preferentially located at 
large distances from the core (i.e., longer baselines), their contribution to the 
column density sensitivity is limited. It should be stressed, of course, that this is 
strictly only the case for \HI emission. For \HI  absorption and continuum 
observations there will be a big difference in resolutions that can be achieved. 
The third aspect is that in order to observe a truly unexplored part of 
resolution-column density parameter space, observing times of $\sim 100$ hours or more
are essential. 

With a survey speed that is five times better than MeerKAT, SKA-Mid will enable detailed and resolved
statistical studies of the low-column density \HI in the local universe. With the same amount of observing 
time as it took to do MHONGOOSE, SKA-Mid can map in the same detail a sample of 150 galaxies, rather
than the 30 that were done with MeerKAT. At the higher resolutions, SKA-Mid is somewhat more sensitive
than MeerKAT, meaning that MHONGOOSE depth can be achieved at $\sim 5''$, rather than the $8''$ of
MHONGOOSE. 
A 100h integration improves the column density sensitivity by 0.5 dex, bringing us in an 
unexplored regime. It also enables observations to MHONGOOSE depth, but at a resolution that is a factor $\sim 2$ higher. Resolving the low-column density gas spatially may be key to further constraining the processes that govern gas accretion. In the following we discuss some of these processes in more detail.

\section{Cold gas outflows and the regulation of mass growth in galaxies}

\subsection{The true mass-loss from outflows requires \HI observations of starburst galaxies}
Feedback, from both star formation and active galactic nuclei, is among the most important mechanisms driving galaxy evolution. Energy and momentum from strong radiation fields, AGNs and supernovae push gas out of the disk plane, reshaping the galaxy and its surroundings. In galaxies like the Milky Way and nearby spirals, this process forms arc-like superbubbles \citep[e.g.][]{MacLow1989}, while in more intense starbursts or strong AGNs it generates galaxy-scale superwinds \citep{Veilleux2005} that transport gas deep into the circumgalactic medium (CGM). Through this gas removal, feedback regulates the mass growth of galaxies. Simulations tell us that without this ejection of gas from galaxies there would be of order 10-100$\times$ greater stellar mass \citep{Pillepich2018}. These winds are therefore critical to setting the properties of the CGM \citep{Tumlinson2017}. While in the local Universe very few galaxies drive superwinds, all evidence suggest that winds were essentially ubiquitous in star forming galaxies at $z\sim1-4$ \citep[reviewed by][]{forster2020}. The current-day mass and overall properties of local Universe galaxies, like the Milky Way, are dependent on the physical mechanisms in those outflows which they experienced in the past. 

Many fundamental questions remain on exactly how this process works \citep{Thompson2024}, with a large amount of uncertainty surrounding the cold gas phases (neutral atomic and molecular) of the wind \citep{Veilleux2020}. Cold gas, the fuel for star formation, plays a critical role in this regulation. If larger amounts of cold gas launch
from a galaxy, then star formation more effectively self-regulates. The relative mass-loss rate from outflows per star formation rate in the galaxy provides a key test to large cosmological simulations \citep{Nelson2019, Pandya2021} as well as high-resolution simulations \citep[e.g.][]{kim2017,Rathjen2021}. These simulations make a clear prediction that more massive galaxies, with more star formation, remove gas less efficiently (lower $\dot{M}_{out}/SFR$). Moreover, \cite{Wright2024} show that the properties of the winds more strongly constrain subgrid physics  in simulations, than do bulk properties like the mass functions of the galaxy. Measurements of the true galactic wind mass-loss rates, therefore, shape our understanding of galaxies. 

More than a decade after ALMA enabled some of the first resolved views of CO in winds \citep{Bolatto2013Nature}, the amount of cold gas in winds remains very uncertain. Because most work has focused on individual, nearby targets very little information is available on the general relationships between outflow mass loss and driving properties such as the star formation rate, mass and general properties of the galaxy. This is needed to constrain theory. To date there are only a handful of galaxies with well-resolved observations of \HI in outflows \citep[e.g.][]{2007AJ....134.1019O,Johnson2012,Martini2018,Mazzilli2025arXiv}.  This challenge has been due to the combined need for sensitivity and spatial resolution from radio interferometers. SKA, therefore, stands to make a leap forward in progress of this subject. 

\begin{figure}
    \centering
    \includegraphics[width=0.99\linewidth]{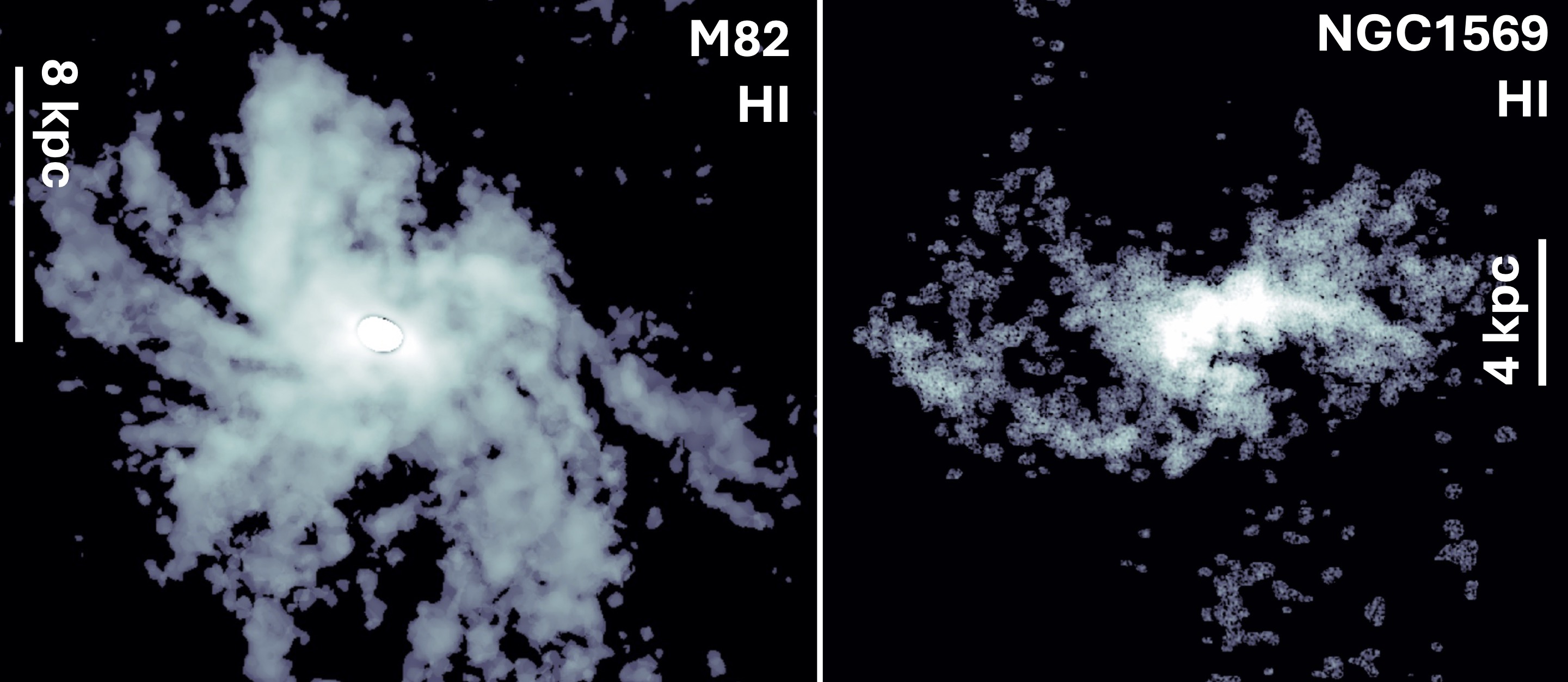}
    \caption{\HI moment 0 maps from VLA observations of the nearest, accessible outflow systems, M~82 (left, \citealp{Martini2018}) and NGC~1569 (right, \citealp{Johnson2012}). In both cases streams of \HI extend away from the disk of the galaxy. These galaxies are understood from optical and X-ray observations to be driving large scale winds \citep[review in][]{Veilleux2005,Veilleux2020}, and the \HI observed perpendicular to the disk is likely significant mass-loss from each galaxy. The nearby proximity of these targets ($\sim$3.2~Mpc for both) is key to mapping \HI with current radio  facilities. }
    \label{fig:himaps} 
\end{figure}

\begin{figure}
    \centering
 \includegraphics[width=0.5\linewidth]{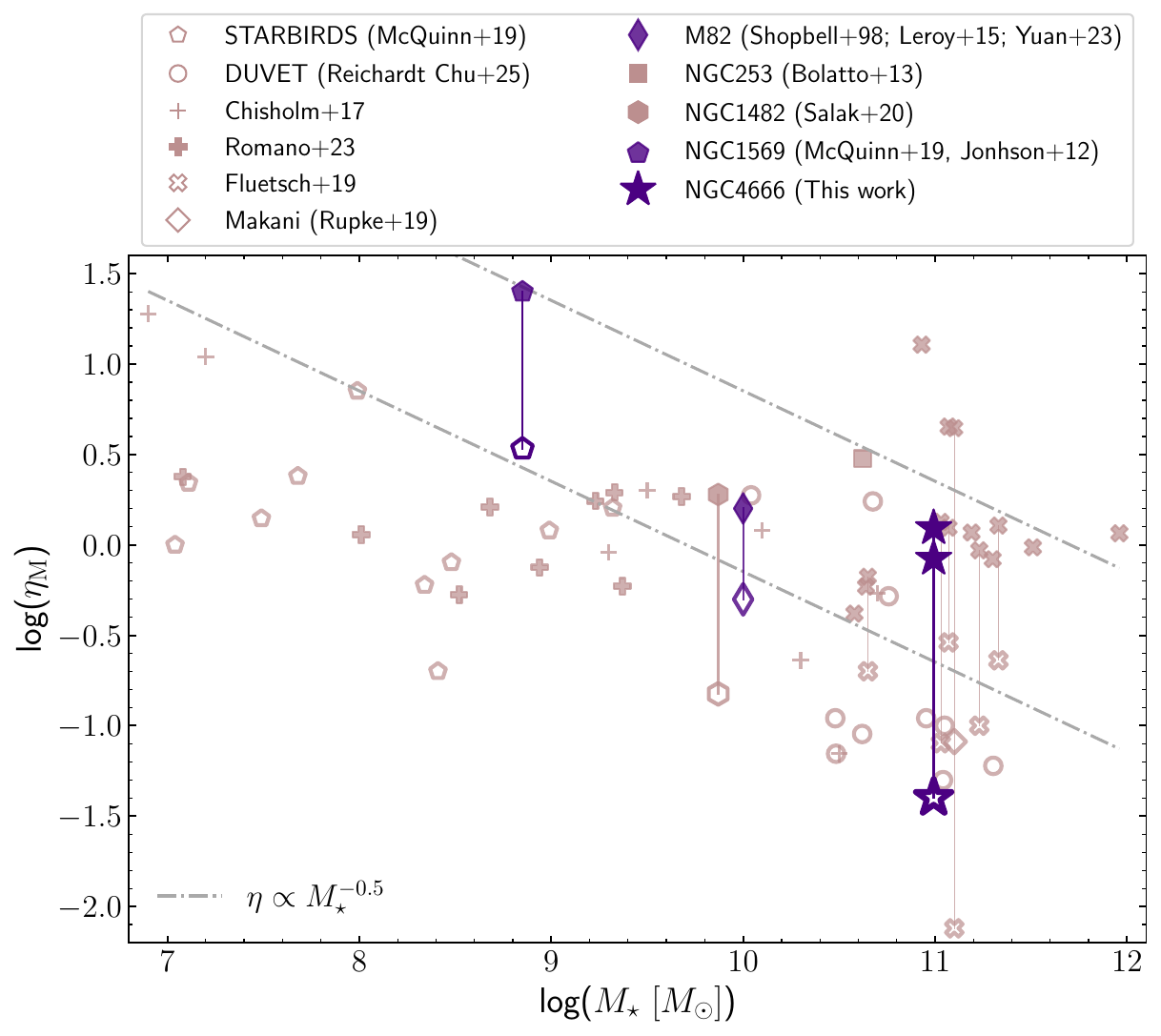}
 \caption{\cite{Mazzilli2025arXiv} compares ionised gas to \HI mass outflow rates in all 3 galaxies that have resolved observations of each. All current data suggests that \HI dominates that mass-outflow rate by a factor of $\sim5-10\times$ over the warm ionised gas. The trend in the relationship of $\dot{M}_{out}/SFR$ with $M_{star}$ is a direct test of galaxy evolution models \citep{Pandya2021,Wright2024}. }
 \label{fig:eta_mstar}
\end{figure}

Figure~\ref{fig:himaps} shows the \HI gas in and around two of the nearest starburst galaxies, M82 and NGC~1569. In both galaxies, recent and ongoing star formation generates sufficient energy to force gas out of the galaxy into the surrounding circumgalactic medium. The images show how \HI gas extends in 4-8~kpc long streams of material from the minor-axis. The \HI substructure takes a U-shape on both sides of the galaxy, consistent with the multiphase model of outflows in which a hot X-ray cone of gas fills in the central part and heats any colder material \citep[e.g.][]{Leroy2015}. Modelling of the M~82 outflow suggests that the cold gas dominates the wind \citep{Xu2023m82, Yuan2023}, with \HI as possibly the largest mass component. This is likewise true in recent analysis of NGC~1569 \citep{Mazzilli2025arXiv}. These starbursts are extremely different, M82 is more massive by a factor of 10 than NGC~1569 and is a highly metal-enriched system, while NGC~1569 is a low mass galaxy with metallicity of $Z\sim0.2~Z_{\odot}$. Given that so few galaxies have \HI, X-ray and ionised gas outflow measurements, the current data suggests  that \HI dominated winds are common, but this requires testing. 

Fig.~\ref{fig:eta_mstar} compares the so-called ``mass-loading" of outlows ($\eta=\dot{M}_{out}/SFR$) to the stellar mass of galaxies for the three galaxies that have measured \HI and ionised gas mass outflow rates. The relationship between the mass-loading and galaxy properties provides one of the strongest tests to galaxy evolution models \citep{Wright2024}. It is therefore critical to determine the full mass leaving the disk. While there are hundreds of measurements of unresolved ionised gas mass-loading in galaxies \citep{forster2020}, and likewise recent integral field spectroscopic programs are now creating resolved observations of tens of outflows \citep[e.g.][]{Reichardt2025}, there are to date very few resolved measurements of \HI outflows. Fig.~\ref{fig:eta_mstar} illustrates that \HI strongly dominates the mass leaving a galaxy. \HI represents the fuel for future star formation; this is different for ionised gas. \HI is, therefore, a more direct tracer of the impact of outflows on the regulation of star formation. Therefore, measurements of \HI mass loss are needed to test theoretical predictions. Models of galaxy evolution predict a decline in the mass-loss efficiency such that $\dot{M}_{out}/SFR\propto M_{star}^{-1/2}$ \citep{Pandya2021}. Currently, the observations used to constrain this are of ionised gas, which do not carry the largest component of mass. We do not know if the relative mass-loss of \HI to ionised gas changes with galaxy mass, star formation rate or other properties such as metallicity. At lower masses, starburst galaxies are more metal-poor and more highly ionised which may impact the details of the multiphase outflows. Moreover, it is not clear how the mass-loading of \HI changes for AGNs and star formation feedback. Because, AGN feedback becomes more important at larger masses, this must be understood to generate full models of galaxy evolution. Surveys of sufficient numbers of outflows are needed to apply this constraint to galaxy evolution. 

\subsection{What SKA can add to mass-loss from outflows}

There is an urgent need for a systematic survey to target outflows in cold gas across a range of galaxy stellar masses. Due to the limits to sensitivity, as well as decreasing brightness of CO for low metallicity galaxies, ALMA is unlikely to achieve this. Because galactic winds are rare, observations must probe beyond the nearest galaxies. Both resolution and sensitivity become critical parameters that require sufficient recovery of outflow flux, with robust seperation of the outflow from the galaxy. SKA-mid AA* will provide kiloparsec-scale resolution at 1.4GHz and reach the required sensitivity in 10hr integration with a 10~arcsec beam. This provides sufficient detections of $\sim2-5\times10^{19}$~cm$^{-2}$, comparable to other outflow measurements \citep[e.g.][]{Martini2018}. Using such data one could robustly probe outflows in galaxies $\sim 50$~Mpc away, sufficient to build samples of $\sim$10 outflow galaxies per 0.5~dex bins in stellar mass \citep{McQuinn2019,Marasco2023, GECKOSpaper}. 

\begin{figure}
    \centering
    \includegraphics[width=0.99\linewidth]{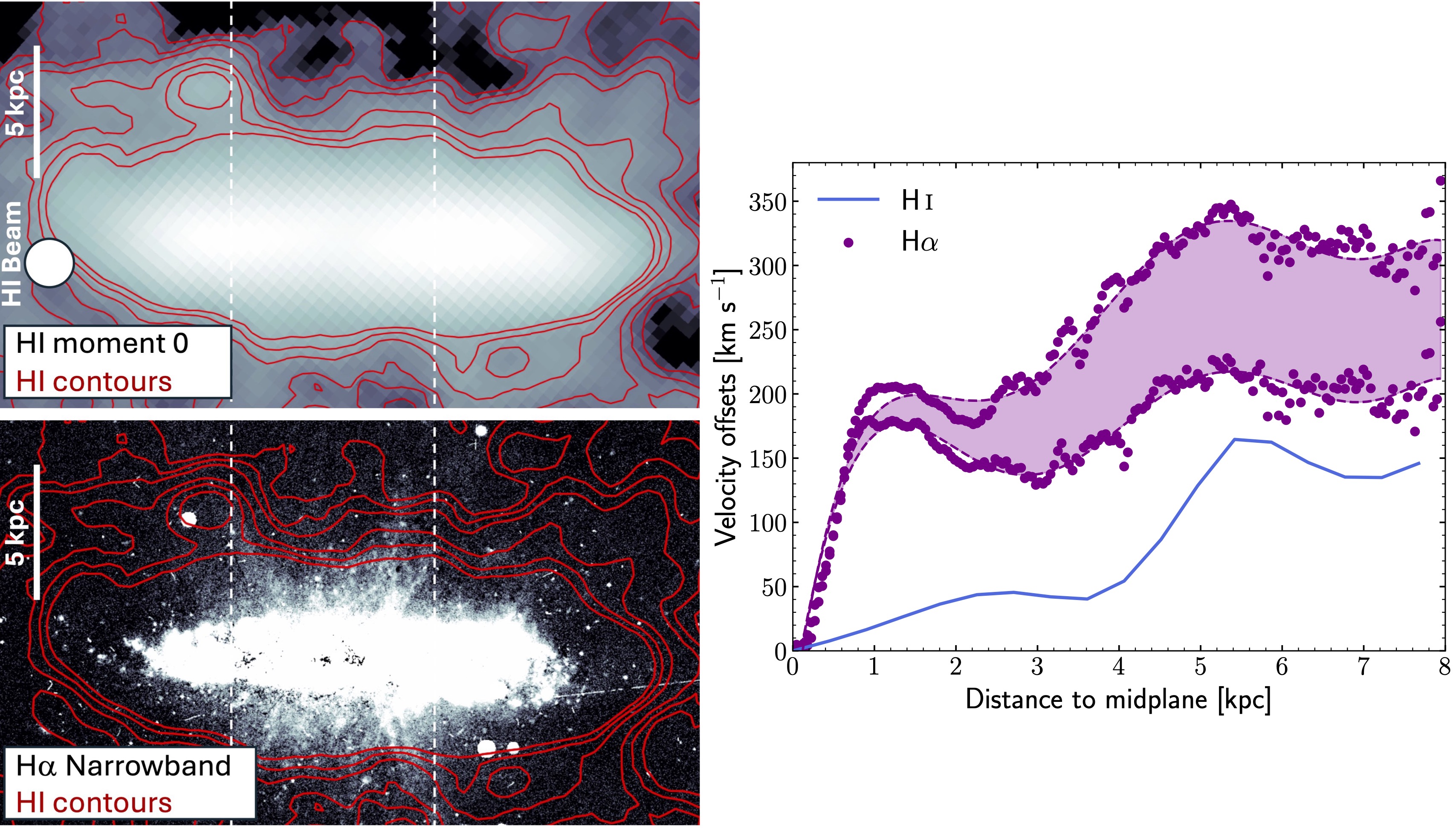}
    \caption{\HI moment 0 map of the outflow in NGC~4666 is compared to H$\alpha$ map of the same target. The biconical outflow is evident at higher spatial resolution, and \HI contours  show a response to the outflow. Using current facilities this can be achieved only on the nearest targets ($\lesssim$15~Mpc), and requires prohibitively long exposures with MeerKAT or VLA.}
    \label{fig:ngc4666}
\end{figure}

At 50~Mpc the 10~arcsec beam has a physical size of $\sim$2.5~kpc. Recently, \cite{Mazzilli2025arXiv} combined ASKAP-WALLABY observations \citep{Lee2022} with VLT/MUSE and narrow-band H$\alpha$ observations of the outflow in nearby galaxy NGC~4666. The galaxy is nearby, $d\sim15.5$~Mpc, and the WALLABY beam corresponds to $\sim$2.5~kpc in this galaxy. Moreover, WALLABY is a relatively shallow, wide area survey. These observations, therefore, demonstrate that even with coarse spatial sampling outflows can be measured at this resolution. As can be seen in Fig.~\ref{fig:ngc4666} the \HI distribution above and below the disk follows the orientation of the H$\alpha$ filaments associated with the outflow. 
 Moreover, the velocity offsets shown in Fig.~\ref{fig:ngc4666} also increase in the outflow region (beyond $\sim 4$ kpc), both for \HI\ and the ionised gas.
This galaxy represents the third strong outflow system where we can  resolve \HI mass-loss due to the wind. This demonstrates the limits of what can be done with \HI observations of very nearby targets. SKA AA* will push this forward significantly. 

\subsection{Galactic fountain regulation of mass growth in Milky Way-like disk galaxies}

The regulation of stellar mass growth by star formation feedback remains a dominant feature of galaxies with more modest star formation rates. Spiral galaxies are understood to maintain an equilibrium in which feedback from young stars and supernovae determines the star formation rate of the galaxy \citep{ostriker2011}. The feedback acts in multiple ways. The injection of energy and momentum in to the ISM generates turbulence, which thickens the disk and reduces star formation. Kinematic observations of both \HI and CO support this picture in local Universe galaxies \citep{Bacchini2020,Girard2021,Lenkic2024}. 

At the same time, energy from supernova also drives gas vertically out of the disk plane, often described as a superbubble \citep{MacLow1989}. 
Superbubbles cool as they expand. The initially hot gas from the disk rains back down as cool neutral gas in a so-called ``galactic fountain" \citep{Shapiro1976}. This process slows star formation in the disk by moving gas into the inner-circumgalactic medium, where it will not form stars. Moreover, cold gas may be entrained in the superbubble as it is launched, which very directly reduces the gas mass available for star formation. 

NGC~891 provides one of the best examples of cold gas in galactic fountains. \cite{Oosterloo2007} show a filament extending $\sim$20~kpc into the circumgalactic medium, upward from the midplane of the disk. The mass of the filament is of order $\sim10^9$~M$_{\odot}$, and is more massive than associated X-ray gas. This suggests that cold gas likewise dominates the ejected mass in lower amplitude galactic fountains, which may be a universal feature of all outflowing gas including both large-scale biconical outflows and smaller galactic fountains. Indeed, the cold gas dominates the outflowing mass even at very low altitudes above the disk plane, which suggests that cold gas entrainment may be a critical feature of galactic fountains \citep[e.g.][]{HodgesKluck2013}.

Recent deep imaging with the FAST 500~m single dish telescope shows vast reservoirs of \HI gas surrounding disk galaxies, including NGC~891 \citep{Yang2025}. These reservoirs strongly suggest a mechanism to maintain them. Moreover, when gas moves out of the plane of disks into the halo, it does not enter into a void of empty space. Cooling of hot halo gas is made much more efficient by the mixing of metal-rich gas from feedback driven winds with existing halo gas \citep{Armillotta2016}, which then will impact the rate of accretion onto the galaxy. The extraplanar gas above a star forming galaxies is, therefore, a mixture of feedback driven winds (and fountains) and accretion onto galaxies (described above). A full characterisation of both processes is necessary to understand the mass growth in galaxies \citep[see review][]{Fraternali2017}. 

Signatures of these processes are found in the kinematics of disk galaxies. ``Lagging" rotation velocity above the plane of the disk has been observed for many years, with the most significant sample \citep{Marasco2019} using the HALOGAS survey. They modelled extraplanar \HI gas in 15 galaxies, and found this thick layer is nearly ubiquitous, and is consistent with models in which this thick layer is fed by the mixture of gas driven to large heights by feedback and gas that is cooling from the halo. Recent work with MeerKAT \citep{Ianjamasimanana2022} shows similar thick layers of \HI around edge-on disk galaxies. 

It is natural to expect this process to vary with the strength of the starburst. Higher gas density in the disk generates larger, higher density star clusters \citep[reviews][]{Krumholz2019,Schinnerer2024}. More massive star clusters naturally generate more supernovae, and thus generate larger superbubbles. What remains very unclear is the amount of cold gas that couples to this feedback, and how far it travels from the disk. High resolution simulations of these processes have significant problems reproducing the cold gas seen in strong winds \citep{Rathjen2021,kim2017}. Direct comparison to systematic studies at star formation rates more similar to Milky Way remain out of reach of current observatories.  

\subsection{What SKA adds to our understanding of galactic fountain driven baryon cycle}

Similar to observations of winds, observing galactic fountains requires a combination of fine spatial resolution and deep sensitivity. While galaxies with galactic fountain-style gas flows are far more common than strong starburst galaxies the resultant superbubbles are smaller, and thus require finer spatial resolution to identify and measure. 

Superbubbles have historically been observed as holes of \HI gas in face-on galaxies \citep[e.g.][]{Boomsma2008}. Because the pressure is decreasing more rapidly in the vertical direction  we expect superbubbles to extend $\sim3-5\times$ further vertically than in the plane of the disk. Statistical analysis of large samples of bubbles in high-spatial resolution JWST imaging of PAH emission \citep{Watkins2023} suggests that sub-kpc spatial resolution is needed for direct imaging of individual bubbles. It is not clear what sensitivity is needed for the extraplanar component of \HI in bubbles. In the disk, \HI holes have a high-surface density, though we expect the z-axis component to be less. The highest resolution achievable would allow for characterization of bubbles with diameter $\sim$200-300~pc in galaxy at a distance 10-15~Mpc. This allows for substructure of the bubble. If fainter observations are necessary, then $\sim1-5\times10^{19}$~cm$^{-2}$ at sub-kpc resolution can still be achieved with 10~hr observations of SKA AA*, for galaxies that are nearer than $\sim15$~Mpc. Theories of feedback make direct predictions for the evolution of a superbubble in comparison to the energy derived from star formation rate. Using SKA \HI observations, one can determine the size, and kinematics of the bubble. This can then be directly compared to SFR of the driver, with radio continuum or another tracer. If bubbles have systematically different sizes and kinematics compared to theory, then modifications to feedback theory are needed. 

\section{Mass accretion from observations of the Milky Way and the Local Group galaxies}

\subsection{Galaxy Evolution and the Gas Depletion Problem}
As already mentioned in the previous sections, galaxies form within dark matter haloes that grow hierarchically in the $\Lambda$CDM cosmology \citep{White1978}, and their primary mass supply is maintained by the smooth accretion of cold gas along the cosmic filaments \citep{Keres2005, Dekel2009}. In the present-day universe, mass accretion onto galaxies is no longer considered the dominant driving mechanism of their growth; instead, mergers are thought to be the primary mode. Galaxy evolution is driven by star formation and its feedback, with the interstellar medium (ISM) serving as the fuel for this process. The currently observed star formation rate in the Milky Way, estimated from infrared satellite observations, is on the order of a few solar masses per year, while the gas mass presently contained is about $10^{9}\ M_{\odot}$. If the star formation rate remains at its current level, it has been pointed out that the gas would be depleted within less than about one billion years. Against this background of the so-called “gas depletion problem,” in this section, we introduce recent studies within the Milky Way on \HI gas clouds falling onto the Galactic disk, and review the impact of mass accretion on star formation in nearby galaxies within the Local Group. We then propose an observational study using \HI observations of external galaxies with the SKA to place constraints on the mode of galaxy evolution in the present-day universe, particularly through quantitative measurements of mass accretion.

\subsection{Observations of Extraplanar Gas in the Milky Way}

\begin{figure}[h]
\centering
\includegraphics[width=0.53\columnwidth]{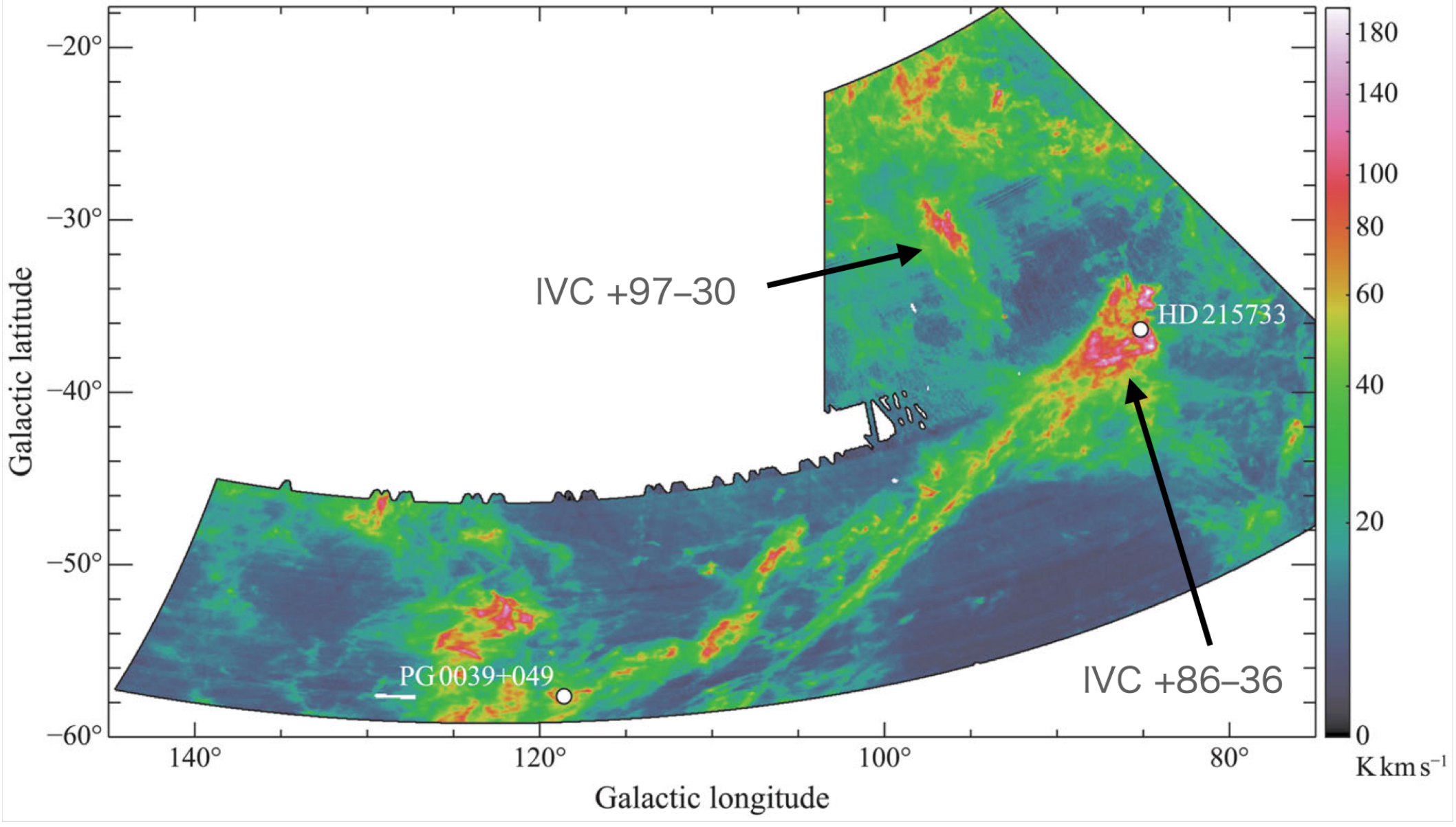}
\includegraphics[width=0.45\columnwidth]{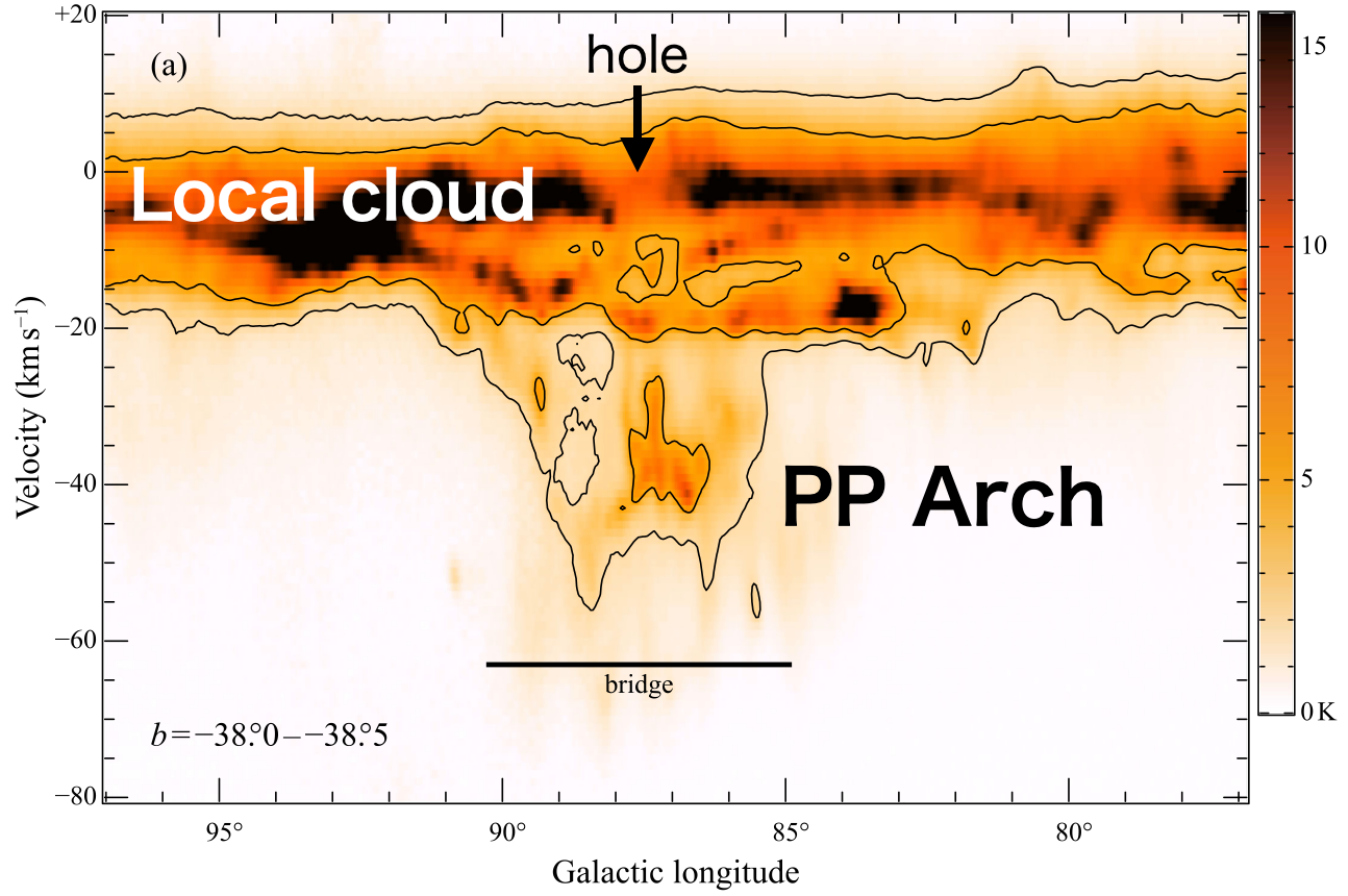}
\caption{Left: Integrated \HI intensity map of the IVC ``PP Arch'' identified from HI4PI data. A characteristic head–tail morphology is evident. Right: Longitude–velocity diagram around the head region of the PP Arch. The IVC shows a blue-shift of about 40 km s$^{-1}$ relative to the nearby local cloud at $\sim0$ km s$^{-1}$, and a diffuse bridge structure connects the two. A cavity is also seen in the local cloud, likely formed as a result of the interaction between them. Modified from \citet{Fukui2021}.}
\label{fig:PP-arch}
\end{figure}

\begin{figure}[h]
\centering
\includegraphics[width=\columnwidth]{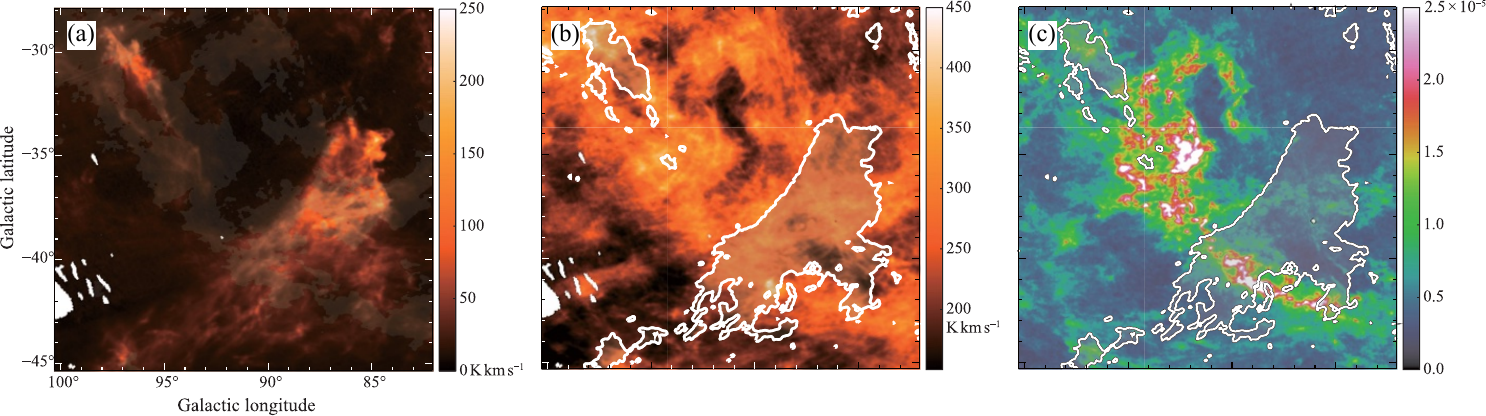}
\caption{(a) Integrated \HI intensity map of the head region of the PP Arch for velocities between $-60$ and $-30$ km s$^{-1}$. Two IVCs, IVC +86--36 and IVC +97--30, are seen. (b) The same region integrated over $-30$ to $+30$ km s$^{-1}$, showing the nearby local cloud. Contours indicate the IVCs from panel (a). (c) Distribution of dust optical depth $\tau_{353}$ in the same region. Contours again trace the IVCs from panel (a). The high-latitude local cloud corresponds to the MBM 53–55 complex, while no significant dust emission is associated with IVC +86--36.
Modified from \citet{Fukui2021}.}
\label{fig:PP-arch dust}
\end{figure}

\HI 21-cm line observations have revealed the presence of High Velocity Clouds (HVCs) and Intermediate Velocity Clouds (IVCs) in the Milky Way. HVCs are clouds moving with velocities deviating by more than 100 km s$^{-1}$ from the Galactic rotation, most of which exhibit negative line-of-sight velocities. In contrast, IVCs have velocities of about 30–90 km s$^{-1}$, and both are thought to be falling toward the Galactic plane. HVCs are generally interpreted as clouds accreting from the Galactic halo onto the disk, while IVCs have traditionally been explained by the ``Galactic fountain'' model, in which gas circulates within the Milky Way \citep{Wakker2004}. This interpretation was primarily based on metallicity measurements from absorption lines: HVCs are composed of low-metallicity gas, whereas IVCs are thought to be metal-rich. However, recent measurements of elemental abundances have shown that the metallicities of IVCs are not as high as previously believed, suggesting that IVCs may have intermediate properties between typical HVCs and local clouds.

\begin{figure}[h]
\centering
\includegraphics[width=\columnwidth]{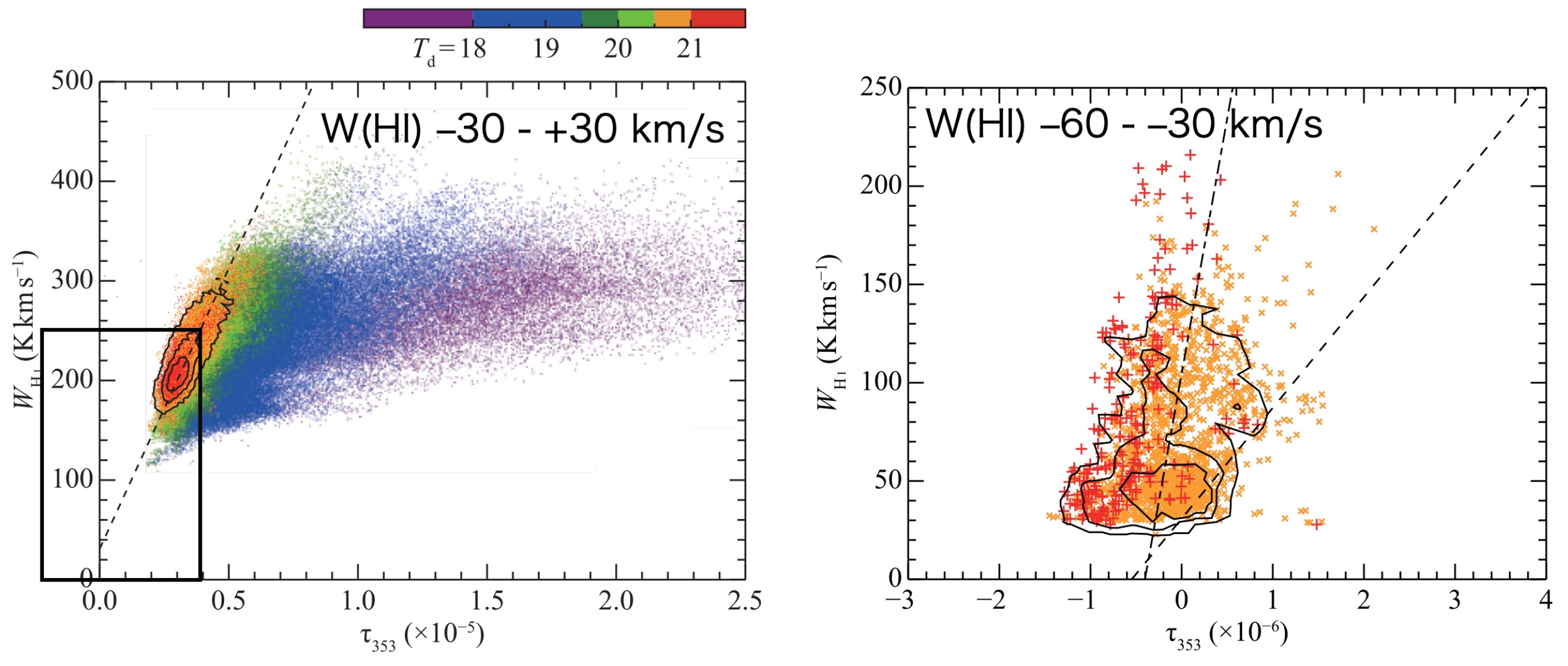}
\caption{Correlation plots between the integrated \HI intensity $W$(HI) and the dust optical depth $\tau_{353}$ for the region shown in Figure~\ref{fig:PP-arch dust}. The left panel corresponds to the velocity range $-30$ to $+30$ km s$^{-1}$ for the local cloud, while the right panel corresponds to the velocity range $-60$ to $-30$ km s$^{-1}$ for the IVC. The square in the left panel indicates the plotting range of the right panel. The dashed line represents the best-fit relation between $W$(\HI) and $\tau_{353}$ derived for optically thin \HI in the local cloud, which has a relatively high dust temperature. The dash–dotted line in the right panel shows the best-fit relation for the IVC velocity range. The slopes of these lines represent the dust-to-gas ratio. Modified from \citet{Fukui2021}.}
\label{fig:dust-gas-ratio}
\end{figure}

These studies have been enabled by all-sky surveys of the \HI 21-cm line and submillimeter dust continuum emission. \citet{Fukui2021} investigated an \HI IVC with a characteristic head–tail morphology, known as the PP Arch (Figure~\ref{fig:PP-arch}). They first found that a remarkable feature of this cloud is its extremely faint dust continuum emission (Figure~\ref{fig:PP-arch dust}). In sharp contrast to the nearby high-latitude cloud complex MBM 53–55, the PP Arch exhibits a dust-to-gas ratio that is estimated to be significantly smaller by a factor of several to an order of magnitude compared with typical local interstellar clouds. The morphology of a dense head followed by a narrow, trailing tail has been well reproduced by numerical simulations \citep{Shelton2022}. These simulations suggest that an \HI cloud falling from a height of $\sim$,a few hundred parsecs above the Galactic disk interacts with the surrounding medium as it approaches the disk, gradually decelerating through ram pressure drag and mixing with the local gas. Indeed, in the PP Arch, clear signatures of interaction are observed between the IVC and the local gas that follows the Galactic rotation.

\begin{figure}[ht]
\centering
\includegraphics[width=\columnwidth]{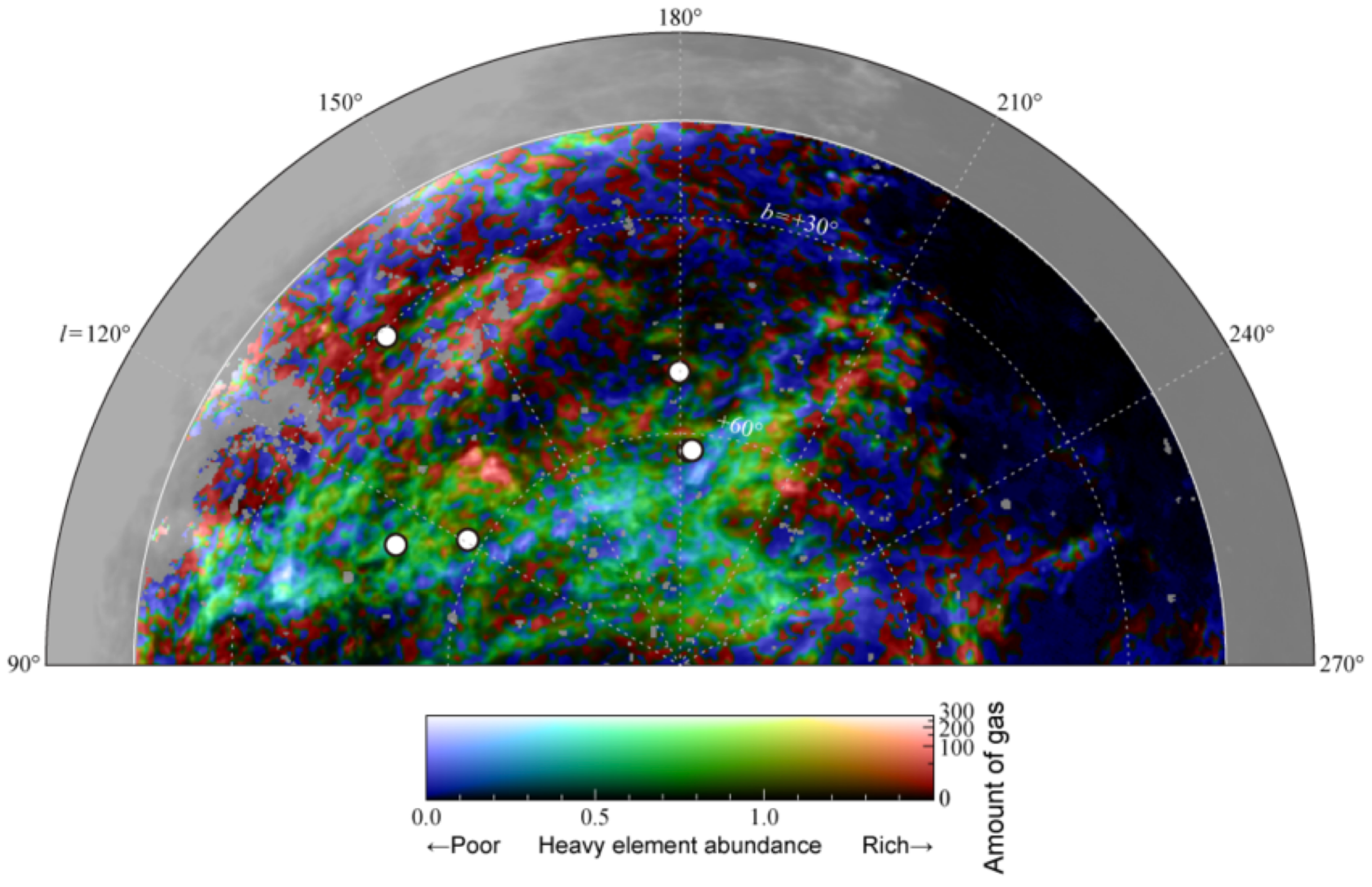}
\caption{Distribution of IVCs in the northern Galactic hemisphere ($b > 0^{\circ}$), covering Galactic longitudes $90^{\circ}$–$270^{\circ}$. Colors indicate the relative metallicity of the interstellar medium, with blue representing metal-poor and red representing metal-rich gas. White circles mark the positions of background sources toward which absorption-line measurements have been obtained. Modified from \citet{Hayakawa2024}.}
\label{fig:quarter-sky}
\end{figure}

The dust-to-gas ratio was measured by dividing the cloud into spatial meshes and calculating the ratio between the \HI integrated intensity and the dust optical depth (Fig.\ \ref{fig:dust-gas-ratio}). The dust optical depth was derived from spectral energy distribution (SED) fitting to multi-band continuum data obtained by the Planck satellite. This approach represented a major improvement over previous methods, which relied solely on absorption-line measurements toward discrete background sources, allowing for the first time a two-dimensional map of the dust-to-gas ratio distribution \citep[see also][]{Fukui2014, Fukui2015, Okamoto2017}. In the case of IVC +86--36, absorption-line spectroscopy toward the background star HD~215733 yielded a subsolar elemental abundance. However, the measurement suffers from large uncertainties due to ionization effects and elemental depletion onto dust grains, as well as the fact that it represents only a single line of sight.

\begin{figure}
    \centering
	\includegraphics[width=6cm]{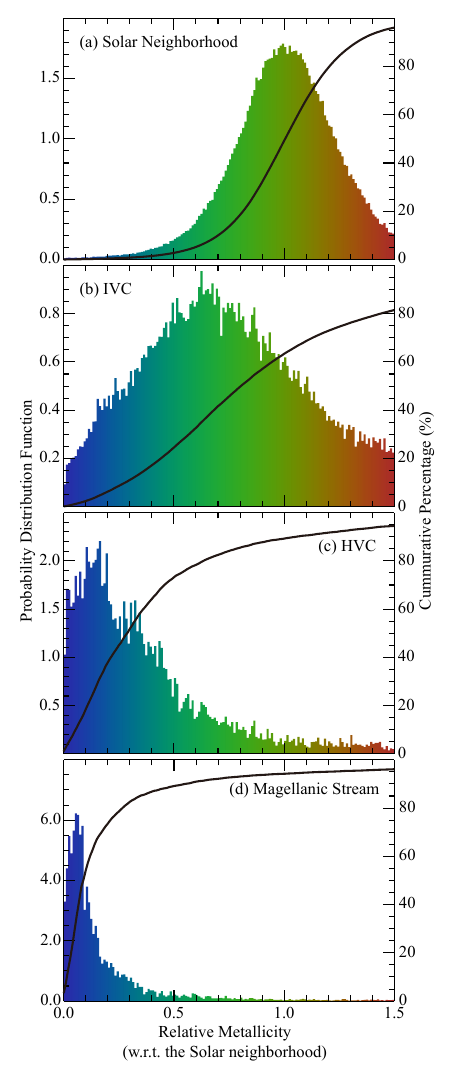}
    \caption{Histograms of the relative dust-to-gas ratio with respect to the average value of local clouds near the Sun, for (a) local clouds, (b) IVCs, (c) HVCs, and (d) the Magellanic Stream. 
    Modified from \citet{Hayakawa2024}.}
    \label{fig:metallicity histograms}
\end{figure}

Building on these results, \citet{Hayakawa2024} extended the same analysis to the entire sky excluding the Galactic plane. Figure~\ref{fig:quarter-sky} shows the distribution of IVCs in one quarter of the northern Galactic hemisphere, revealing that a significant fraction of the clouds exhibit relatively low metallicities. The distribution of IVC metallicities (Fig.\ \ref{fig:metallicity histograms}) demonstrates that they possess intermediate properties between those of HVCs and local clouds. This finding challenges the traditional Galactic fountain scenario—in which material is expelled from the Galactic disk, cools, and subsequently falls back—and instead supports the interpretation proposed from the PP~Arch observations: that IVCs represent decelerated HVCs that have mixed with the more metal-rich interstellar medium near the Galactic plane. If this interpretation is correct, the IVCs currently observed in the Milky Way are expected to lie relatively close to the Galactic disk. Therefore, estimating their total mass across the entire Galaxy remains difficult. Nevertheless, if the inflow rate of low-metallicity hydrogen gas accreting from the Galactic halo is on the order of a few $M_{\odot}$\,yr$^{-1}$, it would be sufficient to sustain the ongoing star formation in the Milky Way. Conversely, if the actual accretion rate is significantly lower, the interstellar gas in the Galaxy would be depleted on a timescale of about one billion years, leading to the eventual cessation of star formation. In this context, the presence of external gas accretion in the current and future Milky Way is of critical importance, and constraining the mass accretion rate provides a key to predicting the Galaxy's future evolution.

In recent years, CO line observations have also revealed molecular clouds that appear to be infalling toward the Galactic disk. \citet{Kohno2025} identified a molecular cloud with a head–tail morphology extending vertically from the Galactic plane. Multi-$J$ CO line observations show that the CO (2–1)/(1–0) line intensity ratio is enhanced in the head region, and LVG analysis indicates that the kinetic temperature ($T_{\rm kin}$) rises above 40 K. Since there is no evidence of associated star formation, this heating is interpreted as shock heating caused by the collision of the infalling cloud with the Galactic disk. Although it is observationally difficult to directly determine the vertical ($Z$-direction) motion of molecular clouds near the Galactic plane, if the motion originates from infalling \HI gas, it implies that both mass and kinetic energy are being supplied from the Galactic halo to the giant molecular clouds (GMCs) in the disk.

Through tidal interactions with the Milky Way, gas has been stripped from the Large and Small Magellanic Clouds, forming the Magellanic Stream (MS), which extends over more than 100 degrees across the sky. The stream is observed as a HVCs system in \HI 21 cm emission \citep{Putman1998, Bruens2005}. Its total mass is estimated to be about $1.6 \times 10^8,M_{\odot}$ \citep{Bruens2005}. In comparison, the total mass of all HVCs is estimated to be $\sim 2.5 \times 10^8,M_{\odot}$ \citep{Wakker2004}, indicating that roughly half of the HVC population originates from the MS. As shown in Figure~\ref{fig:metallicity histograms}, the metallicity of the MS is as low as about 0.1 solar, suggesting an origin in the Small Magellanic Cloud (SMC). By contrast, the metallicity of other HVCs is significantly higher than that of the MS. This implies that the HVCs have undergone substantial metal enrichment, possibly through mixing with interstellar material (ISM) lifted to high altitudes by the galactic fountain, or through interactions with the ISM near the Galactic disk after infall, while still retaining high velocities.

\citet{Fox2014} estimated the mass accretion rate due to the infall of the MS to be about $2,M_{\odot},{\rm yr^{-1}}$. In contrast, \citet{Richter2017} estimated a total accretion rate of $\sim 6.1,M_{\odot},{\rm yr^{-1}}$ when including both neutral and ionized gas from all HVCs, including the MS. This rate is sufficient to sustain, or even exceed, the current star formation rate of the Milky Way. However, considering the present total mass of the HVC system, such gas would be depleted within roughly $10^8$ years. To compensate for this depletion, either additional tidal stripping of gas from nearby satellite galaxies or the cooling and subsequent infall of hot, ionized coronal gas in the Galactic halo would be required.

\subsection{Proposed Observations of Accreting Gas in Nearby Galaxies}

Gas exchange is frequently observed between interacting galaxies in the nearby universe. The most prominent example is the Large and Small Magellanic Clouds, whose connecting structure, the Magellanic Bridge (MB), is thought to have formed as a result of tidal interactions. This process has been successfully reproduced by many numerical simulations \citep[e.g.,][]{Murai1980}, and Gaia observations have revealed stellar motions from the SMC toward the LMC \citep{Schmidt2020}. At the same time, observations have shown that the tidal forces are severely stretching and disrupting the less massive SMC \citep{Nakano2025a, Nakano2025b}. As a consequence of this tidal interaction, it has been suggested that gas accretion from the SMC onto the LMC triggered active star formation observed for example in 30 Doradus through its collision with the LMC disk \citep{Fukui2017, Tsuge2019, Tsuge2024}. Similar mass exchange has been proposed for the M31–M33 system, where active star formation in NGC 604 may have been induced by such an interaction \citep{Tachihara2018}. These phenomena are thought to be driven by high-velocity gas collisions of up to $\sim 100$~km~s$^{-1}$, which efficiently compress and densify the \HI gas. 
Signs of comparable intergalactic interactions have been observed in many other interacting systems. For example, the MHONGOOSE project with MeerKAT detected nine satellite galaxies around the edge-on galaxy UGCA~250, revealing a tail-like \HI structure extending toward one of them \citep{Kurapati2025}. See previous sections for more details about the MHONGOOSE project.

Comparable investigations can also be applied to galaxies that do not exhibit clear evidence of tidal interaction. In particular, for evolved disk galaxies, if cooling of hot ionized coronal gas or intergalactic gas can be directly observed as it condenses and falls toward the disk as HVCs, it would enable estimates of the quasi-steady mass accretion rate onto the galaxy. Observations of edge-on galaxies are especially effective for this purpose. By detecting extraplanar \HI gas located above the galactic plane and separating it from the rotating disk component, one can identify accreting structures. Following the approach used in the Milky Way, correlations between \HI and dust emission should then be examined. Dust emission from such high-altitude extraplanar gas can be effectively traced by submillimeter continuum observations with ALMA. By measuring the metallicity of the extraplanar gas relative to that of the galactic disk, it becomes possible to distinguish whether the HVCs originate from galactic fountain processes or from infalling halo gas, thereby constraining the external mass accretion rate. Conducting such studies for a large sample of nearby galaxies—particularly for evolved disk systems—will enable a statistical investigation of present-day galactic gas dynamics and provide insights into the future evolution of mature galaxies.

Previous observations from projects like VLA THINGS \citep{Walter2008}, ATCA LVHIS \citep{Koribalski2018}, and WSRT HALOGAS (e.g., \citealt{Marasco2019}) have achieved resolutions of about 7$^{\prime\prime}$--30$^{\prime\prime}$. MeerKAT MHONGOOSE has observed nearby galaxies with $~8^{\prime\prime}$ resolution. SKA AA4 will achieve higher spatial resolution and sensitivities.  With a resolution of 6$^{\prime\prime}$ (300 pc at the distance of 10 Mpc), an \HI column density sensitivity of $N_{\rm HI} \sim {\rm a\ few} \times 10^{18}$ cm$^{-2}$ is expected, enabling the detection of \HI clouds of a few $\times 10^{3}\ M_{\odot}$.
The UNGC/LVG database catalogs 869 galaxies within 11 Mpc \citep{Karachentsev2013}. Among these, disk galaxies comprise approximately 20\%. Estimating the number of edge-on galaxies with an axial ratio $\lesssim 0.25$, i.e., $i > 80 ^{\circ}$, yields roughly 20 such galaxies. Galaxies showing no obvious signs of interaction, such as UGC 7321, ESO 274-G001, and IC 5052, are good candidates for isolated galaxies. On the other hand, there are also clear interacting galaxies, such as NGC 3628 with its long tidal tail. Comparative studies of these will investigate the effects of perturbations from nearby galaxies on mass accretion.
Through the statistical studies of many galaxies by SKA AA4, our understanding of mass accretion into evolved disk galaxies will advance dramatically, and additionally will provide clues about the future evolution of the Milky Way Galaxy.

{\bf Acknowledgements}\\
This work has received funding from the European Research Council
(ERC) under the European Union’s Horizon 2020 research and innovation
programme (grant agreement No. 882793 ``MeerGas'').
FMM carried out part of the research activities described in this paper with contribution of the Next Generation EU funds within the National Recovery and Resilience Plan (PNRR), Mission 4 - Education and Research, Component 2 - From Research to Business (M4C2), Investment Line 3.1 - Strengthening and creation of Research Infrastructures, Project IR0000034 – “STILES - Strengthening the Italian Leadership in ELT and SKA”.
LC acknowledges financial support from the French Agence Nationale
de la Recherche ANR.

\bibliographystyle{abbrvnat-maxbibnames4}
\bibliography{chapter} 

\begin{thebibliography}{96}
\providecommand{\natexlab}[1]{#1}
\providecommand{\url}[1]{\texttt{#1}}
\expandafter\ifx\csname urlstyle\endcsname\relax
  \providecommand{\doi}[1]{doi: #1}\else
  \providecommand{\doi}{doi: \begingroup \urlstyle{rm}\Url}\fi

\bibitem[{Armillotta} et~al.(2016){Armillotta}, {Fraternali}, and {Marinacci}]{Armillotta2016}
L.~{Armillotta}, F.~{Fraternali}, and F.~{Marinacci}.
\newblock \emph{\mnras}, 462\penalty0 (4):\penalty0 4157--4170, Nov. 2016.
\newblock \doi{10.1093/mnras/stw1930}.

\bibitem[Auld et~al.(2006)Auld, Minchin, Davies, et~al.]{Auld.2006}
R.~Auld, R.~F. Minchin, J.~I. Davies, et~al.
\newblock \emph{MNRAS}, 371\penalty0 (4):\penalty0 1617, 2006.

\bibitem[{Bacchini} et~al.(2020){Bacchini}, {Fraternali}, {Iorio}, {Pezzulli}, {Marasco}, and {Nipoti}]{Bacchini2020}
C.~{Bacchini} et al.
\newblock \emph{\aap}, 641:\penalty0 A70, Sept. 2020.
\newblock \doi{10.1051/0004-6361/202038223}.

\bibitem[Bagetakos et~al.(2011)Bagetakos, Brinks, Walter, et~al.]{bagetakos}
I.~Bagetakos, E.~Brinks, F.~Walter, et~al.
\newblock \emph{AJ}, 141:\penalty0 23, 2011.

\bibitem[Barnes et~al.(2001)Barnes, Staveley-Smith, de~Blok, et~al.]{Barnes.2001}
D.~G. Barnes, L.~Staveley-Smith, W.~J.~G. de~Blok, et~al.
\newblock \emph{MNRAS}, 322\penalty0 (3):\penalty0 486, 2001.

\bibitem[{Bolatto} et~al.(2013){Bolatto}, {Warren}, {Leroy}, {Walter}, {Veilleux}, {Ostriker}, {Ott}, {Zwaan}, {Fisher}, {Weiss}, {Rosolowsky}, and {Hodge}]{Bolatto2013Nature}
A.~D. {Bolatto} et al.
\newblock \emph{\nat}, 499\penalty0 (7459):\penalty0 450--453, July 2013.
\newblock \doi{10.1038/nature12351}.

\bibitem[{Boomsma} et~al.(2008){Boomsma}, {Oosterloo}, {Fraternali}, {van der Hulst}, and {Sancisi}]{Boomsma2008}
R.~{Boomsma} et al.
\newblock \emph{\aap}, 490\penalty0 (2):\penalty0 555--570, Nov. 2008.
\newblock \doi{10.1051/0004-6361:200810120}.

\bibitem[Bregman(1980)]{bregman}
J.~N. Bregman.
\newblock \emph{ApJ}, 236:\penalty0 577, 1980.

\bibitem[{Br{\"u}ns} et~al.(2005){Br{\"u}ns}, {Kerp}, {Staveley-Smith}, {Mebold}, {Putman}, {Haynes}, {Kalberla}, {Muller}, and {Filipovic}]{Bruens2005}
C.~{Br{\"u}ns} et al.
\newblock \emph{\aap}, 432\penalty0 (1):\penalty0 45--67, Mar. 2005.
\newblock \doi{10.1051/0004-6361:20040321}.

\bibitem[de~Blok et~al.(2024)de~Blok, Healy, Maccagni, et~al.]{deblok24}
W.~J.~G. de~Blok, J.~Healy, F.~M. Maccagni, et~al.
\newblock \emph{A\&A}, 688:\penalty0 A109, 2024.

\bibitem[{Dekel} et~al.(2009){Dekel}, {Birnboim}, {Engel}, {Freundlich}, {Goerdt}, {Mumcuoglu}, {Neistein}, {Pichon}, {Teyssier}, and {Zinger}]{Dekel2009}
A.~{Dekel} et al.
\newblock \emph{\nat}, 457\penalty0 (7228):\penalty0 451--454, Jan. 2009.
\newblock \doi{10.1038/nature07648}.

\bibitem[{F{\"o}rster Schreiber} and {Wuyts}(2020)]{forster2020}
N.~M. {F{\"o}rster Schreiber} and S.~{Wuyts}.
\newblock \emph{\araa}, 58:\penalty0 661--725, Aug. 2020.
\newblock \doi{10.1146/annurev-astro-032620-021910}.

\bibitem[{Fox} et~al.(2014){Fox}, {Wakker}, {Barger}, {Hernandez}, {Richter}, {Lehner}, {Bland-Hawthorn}, {Charlton}, {Westmeier}, {Thom}, {Tumlinson}, {Misawa}, {Howk}, {Haffner}, {Ely}, {Rodriguez-Hidalgo}, and {Kumari}]{Fox2014}
A.~J. {Fox} et al.
\newblock \emph{\apj}, 787\penalty0 (2):\penalty0 147, June 2014.
\newblock \doi{10.1088/0004-637X/787/2/147}.

\bibitem[Fraternali(2013)]{ff13}
F.~Fraternali.
\newblock In \emph{IAU Symposium}, volume 298, 2013.

\bibitem[{Fraternali}(2017)]{Fraternali2017}
F.~{Fraternali}.
\newblock In A.~{Fox} and R.~{Dav{\'e}}, editors, \emph{Gas Accretion onto Galaxies}, volume 430 of \emph{Astrophysics and Space Science Library}, page 323, Jan. 2017.
\newblock \doi{10.1007/978-3-319-52512-9_14}.

\bibitem[{Fukui} et~al.(2014){Fukui}, {Okamoto}, {Kaji}, {Yamamoto}, {Torii}, {Hayakawa}, {Tachihara}, {Dickey}, {Okuda}, {Ohama}, {Kuroda}, and {Kuwahara}]{Fukui2014}
Y.~{Fukui} et al.
\newblock \emph{\apj}, 796\penalty0 (1):\penalty0 59, Nov. 2014.
\newblock \doi{10.1088/0004-637X/796/1/59}.

\bibitem[{Fukui} et~al.(2015){Fukui}, {Torii}, {Onishi}, {Yamamoto}, {Okamoto}, {Hayakawa}, {Tachihara}, and {Sano}]{Fukui2015}
Y.~{Fukui} et al.
\newblock \emph{\apj}, 798\penalty0 (1):\penalty0 6, Jan. 2015.
\newblock \doi{10.1088/0004-637X/798/1/6}.

\bibitem[{Fukui} et~al.(2017){Fukui}, {Tsuge}, {Sano}, {Bekki}, {Yozin}, {Tachihara}, and {Inoue}]{Fukui2017}
Y.~{Fukui} et al.
\newblock \emph{\pasj}, 69\penalty0 (3):\penalty0 L5, June 2017.
\newblock \doi{10.1093/pasj/psx032}.

\bibitem[{Fukui} et~al.(2021){Fukui}, {Koga}, {Maruyama}, {Hayakawa}, {Okamoto}, {Yamamoto}, {Tachihara}, {Shelton}, and {Sasaki}]{Fukui2021}
Y.~{Fukui} et al.
\newblock \emph{\pasj}, 73:\penalty0 S117--S128, Jan. 2021.
\newblock \doi{10.1093/pasj/psy120}.

\bibitem[{Girard} et~al.(2021){Girard}, {Fisher}, {Bolatto}, {Abraham}, {Bassett}, {Glazebrook}, {Herrera-Camus}, {Jim{\'e}nez}, {Lenki{\'c}}, and {Obreschkow}]{Girard2021}
M.~{Girard} et al.
\newblock \emph{\apj}, 909\penalty0 (1):\penalty0 12, Mar. 2021.
\newblock \doi{10.3847/1538-4357/abd5b9}.

\bibitem[{Hayakawa} and {Fukui}(2024)]{Hayakawa2024}
T.~{Hayakawa} and Y.~{Fukui}.
\newblock \emph{\mnras}, 529\penalty0 (1):\penalty0 1--31, Mar. 2024.
\newblock \doi{10.1093/mnras/stae302}.

\bibitem[Haynes et~al.(2018)Haynes, Giovanelli, Kent, et~al.]{Haynes.2018}
M.~P. Haynes, R.~Giovanelli, B.~R. Kent, et~al.
\newblock \emph{ApJ}, 861\penalty0 (1):\penalty0 49, 2018.

\bibitem[Heald et~al.(2011)Heald, J\'ozsa, Serra, et~al.]{heald11}
G.~Heald, G.~J\'ozsa, P.~Serra, et~al.
\newblock \emph{A\&A}, 526:\penalty0 A118, 2011.

\bibitem[{Hodges-Kluck} and {Bregman}(2013)]{HodgesKluck2013}
E.~J. {Hodges-Kluck} and J.~N. {Bregman}.
\newblock \emph{\apj}, 762\penalty0 (1):\penalty0 12, Jan. 2013.
\newblock \doi{10.1088/0004-637X/762/1/12}.

\bibitem[Hunter et~al.(2012)Hunter, Ficut-Vicas, Ashley, et~al.]{hunter}
D.~A. Hunter, D.~Ficut-Vicas, T.~Ashley, et~al.
\newblock \emph{AJ}, 144:\penalty0 134, 2012.

\bibitem[Ianjamasimanana et~al.(2017)Ianjamasimanana, de~Blok, and Heald]{Ianjamasimanana.2017}
R.~Ianjamasimanana, W.~J.~G. de~Blok, and G.~H. Heald.
\newblock \emph{AJ}, 153\penalty0 (5):\penalty0 213, 2017.

\bibitem[{Ianjamasimanana} et~al.(2022){Ianjamasimanana}, {Koribalski}, {J{\'o}zsa}, {Kamphuis}, {de Blok}, {Kleiner}, {Namumba}, {Carignan}, {Dettmar}, {Serra}, {Smirnov}, {Thorat}, {Hugo}, {Ramaila}, {Maina}, {Maccagni}, {Makhathini}, {Andati}, {Moln{\'a}r}, {Perkins}, {Loi}, {Ramatsoku}, and {Atemkeng}]{Ianjamasimanana2022}
R.~{Ianjamasimanana} et al.
\newblock \emph{\mnras}, 513\penalty0 (2):\penalty0 2019--2038, June 2022.
\newblock \doi{10.1093/mnras/stac936}.

\bibitem[Irwin et~al.(2009)Irwin, Hoffman, Spekkens, et~al.]{Irwin.2009}
J.~A. Irwin, G.~L. Hoffman, K.~Spekkens, et~al.
\newblock \emph{ApJ}, 692:\penalty0 1447, 2009.

\bibitem[{Johnson} et~al.(2012){Johnson}, {Hunter}, {Oh}, {Zhang}, {Elmegreen}, {Brinks}, {Tollerud}, and {Herrmann}]{Johnson2012}
M.~{Johnson} et al.
\newblock \emph{\aj}, 144\penalty0 (5):\penalty0 152, Nov. 2012.
\newblock \doi{10.1088/0004-6256/144/5/152}.

\bibitem[{Karachentsev} et~al.(2013){Karachentsev}, {Makarov}, and {Kaisina}]{Karachentsev2013}
I.~D. {Karachentsev}, D.~I. {Makarov}, and E.~I. {Kaisina}.
\newblock \emph{\aj}, 145\penalty0 (4):\penalty0 101, Apr. 2013.
\newblock \doi{10.1088/0004-6256/145/4/101}.

\bibitem[Kere{\v s} et~al.(2005)Kere{\v s}, Katz, Weinberg, and Dav{\'e}]{keres}
D.~Kere{\v s}, N.~Katz, D.~H. Weinberg, and R.~Dav{\'e}.
\newblock \emph{MNRAS}, 363\penalty0 (2), 2005.

\bibitem[{Kere{\v{s}}} et~al.(2005){Kere{\v{s}}}, {Katz}, {Weinberg}, and {Dav{\'e}}]{Keres2005}
D.~{Kere{\v{s}}}, N.~{Katz}, D.~H. {Weinberg}, and R.~{Dav{\'e}}.
\newblock \emph{\mnras}, 363\penalty0 (1):\penalty0 2--28, Oct. 2005.
\newblock \doi{10.1111/j.1365-2966.2005.09451.x}.

\bibitem[{Kim} et~al.(2017){Kim}, {Ostriker}, and {Raileanu}]{kim2017}
C.-G. {Kim}, E.~C. {Ostriker}, and R.~{Raileanu}.
\newblock \emph{\apj}, 834:\penalty0 25, Jan. 2017.
\newblock \doi{10.3847/1538-4357/834/1/25}.

\bibitem[{Kohno} et~al.(2025){Kohno}, {Fukui}, {Hayakawa}, {Doi}, {Yamada}, {Demachi}, {Tokuda}, {Sano}, {Fujita}, {Enokiya}, {Habe}, {Tsuge}, {Nishimura}, {Kobayashi}, {Yamamoto}, and {Tachihara}]{Kohno2025}
M.~{Kohno} et al.
\newblock \emph{arXiv e-prints}, art. arXiv:2510.18399, Oct. 2025.
\newblock \doi{10.48550/arXiv.2510.18399}.

\bibitem[{Koribalski} et~al.(2018){Koribalski}, {Wang}, {Kamphuis}, {Westmeier}, {Staveley-Smith}, {Oh}, {L{\'o}pez-S{\'a}nchez}, {Wong}, {Ott}, {de Blok}, and {Shao}]{Koribalski2018}
B.~S. {Koribalski} et al.
\newblock \emph{\mnras}, 478\penalty0 (2):\penalty0 1611--1648, Aug. 2018.
\newblock \doi{10.1093/mnras/sty479}.

\bibitem[Koribalski et~al.(2018)Koribalski, Wang, Kamphuis, et~al.]{Koribalski.2018}
B.~S. Koribalski, J.~Wang, P.~Kamphuis, et~al.
\newblock \emph{MNRAS}, 478\penalty0 (2):\penalty0 1611, 2018.

\bibitem[Koribalski et~al.(2020)Koribalski, Staveley-Smith, Westmeier, et~al.]{Koribalski.2020}
B.~S. Koribalski, L.~Staveley-Smith, T.~Westmeier, et~al.
\newblock \emph{Ap\&SS}, 365\penalty0 (7):\penalty0 118, 2020.

\bibitem[{Krumholz} et~al.(2019){Krumholz}, {McKee}, and {Bland-Hawthorn}]{Krumholz2019}
M.~R. {Krumholz}, C.~F. {McKee}, and J.~{Bland-Hawthorn}.
\newblock \emph{\araa}, 57:\penalty0 227--303, Aug. 2019.
\newblock \doi{10.1146/annurev-astro-091918-104430}.

\bibitem[{Kurapati} et~al.(2025){Kurapati}, {Pisano}, {de Blok}, {Kamphuis}, {Zabel}, {de Villiers}, {Healy}, {Maccagni}, {Kleiner}, {Adams}, {Amram}, {Athanassoula}, {Bigiel}, {Bosma}, {Brinks}, {Chemin}, {Combes}, {Dettmar}, {J{\'o}zsa}, {Koribalski}, {Marasco}, {Meurer}, {Mogotsi}, {Mohapatra}, {Rajohnson}, {Schinnerer}, {Sorgho}, {Spekkens}, {Verdes-Montenegro}, {Veronese}, and {Walter}]{Kurapati2025}
S.~{Kurapati} et al.
\newblock \emph{\mnras}, 538\penalty0 (2):\penalty0 1272--1287, Apr. 2025.
\newblock \doi{10.1093/mnras/staf387}.

\bibitem[{Lee} et~al.(2022){Lee}, {Wang}, {Chung}, {Ho}, {Wang}, {Michiyama}, {Molina}, {Kim}, {Shao}, {Kilborn}, {Wang}, {Lin}, {Kim}, {Catinella}, {Cortese}, {Deg}, {Denes}, {Elagali}, {For}, {Kleiner}, {Koribalski}, {Lee-Waddell}, {Rhee}, {Spekkens}, {Westmeier}, {Wong}, {Bigiel}, {Bosma}, {Holwerda}, {van der Hulst}, {Roychowdhury}, {Verdes-Montenegro}, and {Zwaan}]{Lee2022}
B.~{Lee} et al.
\newblock \emph{\apjs}, 262\penalty0 (1):\penalty0 31, Sept. 2022.
\newblock \doi{10.3847/1538-4365/ac7eba}.

\bibitem[{Lenki{\'c}} et~al.(2024){Lenki{\'c}}, {Fisher}, {Bolatto}, {Teuben}, {Levy}, {Sun}, {Herrera-Camus}, {Glazebrook}, {Obreschkow}, and {Abraham}]{Lenkic2024}
L.~{Lenki{\'c}} et al.
\newblock \emph{\apj}, 976\penalty0 (1):\penalty0 88, Nov. 2024.
\newblock \doi{10.3847/1538-4357/ad758c}.

\bibitem[{Leroy} et~al.(2015){Leroy}, {Walter}, {Martini}, {Roussel}, {Sandstrom}, {Ott}, {Weiss}, {Bolatto}, {Schuster}, and {Dessauges-Zavadsky}]{Leroy2015}
A.~K. {Leroy} et al.
\newblock \emph{\apj}, 814\penalty0 (2):\penalty0 83, Dec. 2015.
\newblock \doi{10.1088/0004-637X/814/2/83}.

\bibitem[{Mac Low} et~al.(1989){Mac Low}, {McCray}, and {Norman}]{MacLow1989}
M.-M. {Mac Low}, R.~{McCray}, and M.~L. {Norman}.
\newblock \emph{\apj}, 337:\penalty0 141, Feb. 1989.
\newblock \doi{10.1086/167094}.

\bibitem[Maccagni and de~Blok(2024)]{maccagni24}
F.~M. Maccagni and W.~J.~G. de~Blok.
\newblock \emph{arXiv preprint}, 2024.
\newblock \doi{10.48550/arXiv.2407.03166}.

\bibitem[{Marasco} et~al.(2019){Marasco}, {Fraternali}, {Heald}, {de Blok}, {Oosterloo}, {Kamphuis}, {J{\'o}zsa}, {Vargas}, {Winkel}, {Walterbos}, {Dettmar}, and {Juẗte}]{Marasco2019}
A.~{Marasco} et al.
\newblock \emph{\aap}, 631:\penalty0 A50, Nov. 2019.
\newblock \doi{10.1051/0004-6361/201936338}.

\bibitem[{Marasco} et~al.(2023){Marasco}, {Belfiore}, {Cresci}, {Lelli}, {Venturi}, {Hunt}, {Concas}, {Marconi}, {Mannucci}, {Mingozzi}, {McLeod}, {Kumari}, {Carniani}, {Vanzi}, and {Ginolfi}]{Marasco2023}
A.~{Marasco} et al.
\newblock \emph{\aap}, 670:\penalty0 A92, Feb. 2023.
\newblock \doi{10.1051/0004-6361/202244895}.

\bibitem[Marasco et~al.(2025)Marasco, de~Blok, Maccagni, et~al.]{marasco25}
A.~Marasco, W.~J.~G. de~Blok, F.~M. Maccagni, et~al.
\newblock \emph{A\&A}, 697:\penalty0 A86, 2025.

\bibitem[{Martini} et~al.(2018){Martini}, {Leroy}, {Mangum}, {Bolatto}, {Keating}, {Sandstrom}, and {Walter}]{Martini2018}
P.~{Martini} et al.
\newblock \emph{\apj}, 856\penalty0 (1):\penalty0 61, Mar. 2018.
\newblock \doi{10.3847/1538-4357/aab08e}.

\bibitem[{Mazzilli Ciraulo} et~al.(2025){Mazzilli Ciraulo}, {Fisher}, {Elliott}, {Fraser-McKelvie}, {Hayden}, {Martig}, {van de Sande}, {Battisti}, {Bland-Hawthorn}, {Bolatto}, {Brown}, {Catinella}, {Combes}, {Cortese}, {Davis}, {Emsellem}, {Gadotti}, {Lagos}, {Lin}, {Marasco}, {Peng}, {Pinna}, {Puzia}, {Silva-Lima}, {Valenzuela}, {van de Ven}, and {Wang}]{Mazzilli2025arXiv}
B.~{Mazzilli Ciraulo} et al.
\newblock \emph{arXiv e-prints}, art. arXiv:2509.17560, Sept. 2025.
\newblock \doi{10.48550/arXiv.2509.17560}.

\bibitem[{McQuinn} et~al.(2019){McQuinn}, {van Zee}, and {Skillman}]{McQuinn2019}
K.~B.~W. {McQuinn}, L.~{van Zee}, and E.~D. {Skillman}.
\newblock \emph{\apj}, 886\penalty0 (1):\penalty0 74, Nov. 2019.
\newblock \doi{10.3847/1538-4357/ab4c37}.

\bibitem[Meyer et~al.(2004)Meyer, Zwaan, Webster, et~al.]{Meyer.2004}
M.~J. Meyer, M.~A. Zwaan, R.~L. Webster, et~al.
\newblock \emph{MNRAS}, 350\penalty0 (4):\penalty0 1195, 2004.

\bibitem[{Murai} and {Fujimoto}(1980)]{Murai1980}
T.~{Murai} and M.~{Fujimoto}.
\newblock \emph{\pasj}, 32\penalty0 (4):\penalty0 581--603, Dec. 1980.
\newblock \doi{10.1093/pasj/32.4.581}.

\bibitem[{Nakano} and {Tachihara}(2025)]{Nakano2025b}
S.~{Nakano} and K.~{Tachihara}.
\newblock \emph{\apjl}, 985\penalty0 (1):\penalty0 L5, May 2025.
\newblock \doi{10.3847/2041-8213/adce0b}.

\bibitem[{Nakano} et~al.(2025){Nakano}, {Tachihara}, and {Tamashiro}]{Nakano2025a}
S.~{Nakano}, K.~{Tachihara}, and M.~{Tamashiro}.
\newblock \emph{\apjs}, 277\penalty0 (2):\penalty0 62, Apr. 2025.
\newblock \doi{10.3847/1538-4365/adb8de}.

\bibitem[{Nelson} et~al.(2019){Nelson}, {Pillepich}, {Springel}, {Pakmor}, {Weinberger}, {Genel}, {Torrey}, {Vogelsberger}, {Marinacci}, and {Hernquist}]{Nelson2019}
D.~{Nelson} et al.
\newblock \emph{\mnras}, 490\penalty0 (3):\penalty0 3234--3261, Dec. 2019.
\newblock \doi{10.1093/mnras/stz2306}.

\bibitem[Norman and Ikeuchi(1989)]{norman}
C.~A. Norman and S.~Ikeuchi.
\newblock \emph{ApJ}, 345:\penalty0 372, 1989.

\bibitem[{Okamoto} et~al.(2017){Okamoto}, {Yamamoto}, {Tachihara}, {Hayakawa}, {Hayashi}, and {Fukui}]{Okamoto2017}
R.~{Okamoto} et al.
\newblock \emph{\apj}, 838\penalty0 (2):\penalty0 132, Apr. 2017.
\newblock \doi{10.3847/1538-4357/aa6747}.

\bibitem[{Oosterloo} et~al.(2007{\natexlab{a}}){Oosterloo}, {Fraternali}, and {Sancisi}]{2007AJ....134.1019O}
T.~{Oosterloo}, F.~{Fraternali}, and R.~{Sancisi}.
\newblock \emph{\aj}, 134\penalty0 (3):\penalty0 1019, Sept. 2007{\natexlab{a}}.
\newblock \doi{10.1086/520332}.

\bibitem[{Oosterloo} et~al.(2007{\natexlab{b}}){Oosterloo}, {Fraternali}, and {Sancisi}]{Oosterloo2007}
T.~{Oosterloo}, F.~{Fraternali}, and R.~{Sancisi}.
\newblock \emph{\aj}, 134\penalty0 (3):\penalty0 1019, Sept. 2007{\natexlab{b}}.
\newblock \doi{10.1086/520332}.

\bibitem[{Ostriker} and {Shetty}(2011)]{ostriker2011}
E.~C. {Ostriker} and R.~{Shetty}.
\newblock \emph{\apj}, 731:\penalty0 41, Apr. 2011.
\newblock \doi{10.1088/0004-637X/731/1/41}.

\bibitem[{Pandya} et~al.(2021){Pandya}, {Fielding}, {Angl{\'e}s-Alc{\'a}zar}, {Somerville}, {Bryan}, {Hayward}, {Stern}, {Kim}, {Quataert}, {Forbes}, {Faucher-Gigu{\`e}re}, {Feldmann}, {Hafen}, {Hopkins}, {Kere{\v{s}}}, {Murray}, and {Wetzel}]{Pandya2021}
V.~{Pandya} et al.
\newblock \emph{\mnras}, 508\penalty0 (2):\penalty0 2979--3008, Dec. 2021.
\newblock \doi{10.1093/mnras/stab2714}.

\bibitem[{Pillepich} et~al.(2018){Pillepich}, {Springel}, {Nelson}, {Genel}, {Naiman}, {Pakmor}, {Hernquist}, {Torrey}, {Vogelsberger}, {Weinberger}, and {Marinacci}]{Pillepich2018}
A.~{Pillepich} et al.
\newblock \emph{\mnras}, 473\penalty0 (3):\penalty0 4077--4106, Jan. 2018.
\newblock \doi{10.1093/mnras/stx2656}.

\bibitem[Pingel et~al.(2018)Pingel, Pisano, Heald, et~al.]{Pingel.2018}
N.~M. Pingel, D.~J. Pisano, G.~Heald, et~al.
\newblock \emph{ApJ}, 865\penalty0 (1):\penalty0 36, 2018.

\bibitem[{Putman} et~al.(1998){Putman}, {Gibson}, {Staveley-Smith}, {Banks}, {Barnes}, {Bhatal}, {Disney}, {Ekers}, {Freeman}, {Haynes}, {Henning}, {Jerjen}, {Kilborn}, {Koribalski}, {Knezek}, {Malin}, {Mould}, {Oosterloo}, {Price}, {Ryder}, {Sadler}, {Stewart}, {Stootman}, {Vaile}, {Webster}, and {Wright}]{Putman1998}
M.~E. {Putman} et al.
\newblock \emph{\nat}, 394\penalty0 (6695):\penalty0 752--754, Aug. 1998.
\newblock \doi{10.1038/29466}.

\bibitem[Putman et~al.(2012)Putman, Peek, and Joung]{putman}
M.~E. Putman, J.~E.~G. Peek, and M.~R. Joung.
\newblock \emph{ARA\&A}, 50:\penalty0 491, 2012.

\bibitem[{Rathjen} et~al.(2021){Rathjen}, {Naab}, {Girichidis}, {Walch}, {W{\"u}nsch}, {Dinnbier}, {Seifried}, {Klessen}, and {Glover}]{Rathjen2021}
T.-E. {Rathjen} et al.
\newblock \emph{\mnras}, 504\penalty0 (1):\penalty0 1039--1061, June 2021.
\newblock \doi{10.1093/mnras/stab900}.

\bibitem[{Reichardt Chu} et~al.(2025){Reichardt Chu}, {Fisher}, {Chisholm}, {Berg}, {Bolatto}, {Cameron}, {Fielding}, {Herrera-Camus}, {Kacprzak}, {Li}, {McLeod}, {McPherson}, {Nielsen}, {Rickards Vaught}, {Ridolfo}, and {Sandstrom}]{Reichardt2025}
B.~{Reichardt Chu} et al.
\newblock \emph{\mnras}, 536\penalty0 (2):\penalty0 1799--1821, Jan. 2025.
\newblock \doi{10.1093/mnras/stae2705}.

\bibitem[{Richter} et~al.(2017){Richter}, {Nuza}, {Fox}, {Wakker}, {Lehner}, {Ben Bekhti}, {Fechner}, {Wendt}, {Howk}, {Muzahid}, {Ganguly}, and {Charlton}]{Richter2017}
P.~{Richter} et al.
\newblock \emph{\aap}, 607:\penalty0 A48, Nov. 2017.
\newblock \doi{10.1051/0004-6361/201630081}.

\bibitem[Sancisi et~al.(2008)Sancisi, Fraternali, Oosterloo, and van~der Hulst]{sancisi}
R.~Sancisi, F.~Fraternali, T.~Oosterloo, and T.~van~der Hulst.
\newblock \emph{A\&A Review}, 15:\penalty0 189, 2008.

\bibitem[Sardone et~al.(2021)Sardone, Pisano, Pingel, et~al.]{Sardone.2021}
A.~Sardone, D.~J. Pisano, N.~M. Pingel, et~al.
\newblock \emph{ApJ}, 910\penalty0 (1):\penalty0 69, 2021.

\bibitem[{Schinnerer} and {Leroy}(2024)]{Schinnerer2024}
E.~{Schinnerer} and A.~K. {Leroy}.
\newblock \emph{\araa}, 62\penalty0 (1):\penalty0 369--436, Sept. 2024.
\newblock \doi{10.1146/annurev-astro-071221-052651}.

\bibitem[{Schmidt} et~al.(2020){Schmidt}, {Cioni}, {Niederhofer}, {Bekki}, {Bell}, {de Grijs}, {Diaz}, {El Youssoufi}, {Emerson}, {Groenewegen}, {Ivanov}, {Matijevic}, {Oliveira}, {Petr-Gotzens}, {Queiroz}, {Ripepi}, and {van Loon}]{Schmidt2020}
T.~{Schmidt} et al.
\newblock \emph{\aap}, 641:\penalty0 A134, Sept. 2020.
\newblock \doi{10.1051/0004-6361/202037478}.

\bibitem[{Shapiro} and {Field}(1976)]{Shapiro1976}
P.~R. {Shapiro} and G.~B. {Field}.
\newblock \emph{\apj}, 205:\penalty0 762--765, May 1976.
\newblock \doi{10.1086/154332}.

\bibitem[Shapiro and Field(1976)]{shapiro}
P.~R. Shapiro and G.~B. Field.
\newblock \emph{ApJ}, 205:\penalty0 762, 1976.

\bibitem[{Shelton} et~al.(2022){Shelton}, {Williams}, {Parker}, {Galyardt}, {Fukui}, and {Tachihara}]{Shelton2022}
R.~L. {Shelton} et al.
\newblock \emph{\apj}, 925\penalty0 (2):\penalty0 190, Feb. 2022.
\newblock \doi{10.3847/1538-4357/ac39a4}.

\bibitem[Sorgho et~al.(2019)Sorgho, Carignan, Pisano, et~al.]{Sorgho.2019}
A.~Sorgho, C.~Carignan, D.~J. Pisano, et~al.
\newblock \emph{MNRAS}, 482\penalty0 (1):\penalty0 1248, 2019.

\bibitem[{Tachihara} et~al.(2018){Tachihara}, {Gratier}, {Sano}, {Tsuge}, {Miura}, {Muraoka}, and {Fukui}]{Tachihara2018}
K.~{Tachihara} et al.
\newblock \emph{\pasj}, 70:\penalty0 S52, May 2018.
\newblock \doi{10.1093/pasj/psy020}.

\bibitem[{Thompson} and {Heckman}(2024)]{Thompson2024}
T.~A. {Thompson} and T.~M. {Heckman}.
\newblock \emph{\araa}, 62\penalty0 (1):\penalty0 529--591, Sept. 2024.
\newblock \doi{10.1146/annurev-astro-041224-011924}.

\bibitem[{Tsuge} et~al.(2019){Tsuge}, {Sano}, {Tachihara}, {Yozin}, {Bekki}, {Inoue}, {Mizuno}, {Kawamura}, {Onishi}, and {Fukui}]{Tsuge2019}
K.~{Tsuge} et al.
\newblock \emph{\apj}, 871\penalty0 (1):\penalty0 44, Jan. 2019.
\newblock \doi{10.3847/1538-4357/aaf4fb}.

\bibitem[{Tsuge} et~al.(2024){Tsuge}, {Sano}, {Tachihara}, {Bekki}, {Tokuda}, {Inoue}, {Mizuno}, {Kawamura}, {Onishi}, and {Fukui}]{Tsuge2024}
K.~{Tsuge} et al.
\newblock \emph{\pasj}, 76\penalty0 (4):\penalty0 589--615, Aug. 2024.
\newblock \doi{10.1093/pasj/psae035}.

\bibitem[{Tumlinson} et~al.(2017){Tumlinson}, {Peeples}, and {Werk}]{Tumlinson2017}
J.~{Tumlinson}, M.~S. {Peeples}, and J.~K. {Werk}.
\newblock \emph{\araa}, 55\penalty0 (1):\penalty0 389--432, Aug. 2017.
\newblock \doi{10.1146/annurev-astro-091916-055240}.

\bibitem[{van de Sande} et~al.(2023){van de Sande}, {Fraser-McKelvie}, {Fisher}, {Martig}, {Hayden}, and {the GECKOS Survey collaboration}]{GECKOSpaper}
J.~{van de Sande} et al.
\newblock \emph{arXiv e-prints}, art. arXiv:2306.00059, May 2023.
\newblock \doi{10.48550/arXiv.2306.00059}.

\bibitem[van~der Hulst et~al.(2001)van~der Hulst, van Albada, and Sancisi]{whisp}
J.~M. van~der Hulst, T.~S. van Albada, and R.~Sancisi.
\newblock In \emph{ASP Conference Series}, volume 240, page 451, 2001.

\bibitem[{Veilleux} et~al.(2005){Veilleux}, {Cecil}, and {Bland-Hawthorn}]{Veilleux2005}
S.~{Veilleux}, G.~{Cecil}, and J.~{Bland-Hawthorn}.
\newblock \emph{\araa}, 43\penalty0 (1):\penalty0 769--826, Sept. 2005.
\newblock \doi{10.1146/annurev.astro.43.072103.150610}.

\bibitem[{Veilleux} et~al.(2020){Veilleux}, {Maiolino}, {Bolatto}, and {Aalto}]{Veilleux2020}
S.~{Veilleux}, R.~{Maiolino}, A.~D. {Bolatto}, and S.~{Aalto}.
\newblock \emph{\aapr}, 28\penalty0 (1):\penalty0 2, Apr. 2020.
\newblock \doi{10.1007/s00159-019-0121-9}.

\bibitem[{Wakker}(2004)]{Wakker2004}
B.~P. {Wakker}.
\newblock In H.~{van Woerden}, B.~P. {Wakker}, U.~J. {Schwarz}, and K.~S. {de Boer}, editors, \emph{High Velocity Clouds}, volume 312 of \emph{Astrophysics and Space Science Library}, page~25, Jan. 2004.
\newblock \doi{10.1007/1-4020-2579-3_2}.

\bibitem[{Walter} et~al.(2008){Walter}, {Brinks}, {de Blok}, {Bigiel}, {Kennicutt}, {Thornley}, and {Leroy}]{Walter2008}
F.~{Walter} et al.
\newblock \emph{\aj}, 136\penalty0 (6):\penalty0 2563--2647, Dec. 2008.
\newblock \doi{10.1088/0004-6256/136/6/2563}.

\bibitem[Walter et~al.(2008)Walter, Brinks, de~Blok, et~al.]{walter08}
F.~Walter, E.~Brinks, W.~J.~G. de~Blok, et~al.
\newblock \emph{AJ}, 136:\penalty0 2563, 2008.

\bibitem[{Wang} et~al.(2024){Wang}, {Lin}, {Yang}, {Staveley-Smith}, {Walter}, {Wang}, {Wang}, {Battisti}, {Catinella}, {Chen}, {Cortese}, {Fisher}, {Ho}, {Ji}, {Jiang}, {Kauffmann}, {Kong}, {Liu}, {Shao}, {Wang}, {Wang}, and {Wang}]{feasts}
J.~{Wang} et al.
\newblock \emph{\apj}, 968\penalty0 (1):\penalty0 48, June 2024.
\newblock \doi{10.3847/1538-4357/ad3e61}.

\bibitem[{Watkins} et~al.(2023){Watkins}, {Kreckel}, {Groves}, {Glover}, {Whitmore}, {Leroy}, {Schinnerer}, {Meidt}, {Egorov}, {Barnes}, {Lee}, {Bigiel}, {Boquien}, {Chandar}, {Chevance}, {Dale}, {Grasha}, {Klessen}, {Kruijssen}, {Larson}, {Li}, {M{\'e}ndez-Delgado}, {Pessa}, {Saito}, {Sanchez-Blazquez}, {Sarbadhicary}, {Scheuermann}, {Thilker}, and {Williams}]{Watkins2023}
E.~J. {Watkins} et al.
\newblock \emph{\aap}, 676:\penalty0 A67, Aug. 2023.
\newblock \doi{10.1051/0004-6361/202346075}.

\bibitem[{White} and {Rees}(1978)]{White1978}
S.~D.~M. {White} and M.~J. {Rees}.
\newblock \emph{\mnras}, 183:\penalty0 341--358, May 1978.
\newblock \doi{10.1093/mnras/183.3.341}.

\bibitem[Wolfe et~al.(2016)Wolfe, Lockman, and Pisano]{Wolfe.2016}
S.~A. Wolfe, F.~J. Lockman, and D.~J. Pisano.
\newblock \emph{ApJ}, 816\penalty0 (2):\penalty0 81, 2016.

\bibitem[{Wright} et~al.(2024){Wright}, {Somerville}, {Lagos}, {Schaller}, {Dav{\'e}}, {Angl{\'e}s-Alc{\'a}zar}, and {Genel}]{Wright2024}
R.~J. {Wright} et al.
\newblock \emph{\mnras}, 532\penalty0 (3):\penalty0 3417--3440, Aug. 2024.
\newblock \doi{10.1093/mnras/stae1688}.

\bibitem[{Xu} et~al.(2023){Xu}, {Heckman}, {Yoshida}, {Henry}, and {Ohyama}]{Xu2023m82}
X.~{Xu} et al.
\newblock \emph{\apj}, 956\penalty0 (2):\penalty0 142, Oct. 2023.
\newblock \doi{10.3847/1538-4357/acfa71}.

\bibitem[{Yang} et~al.(2025){Yang}, {Wang}, {Qu}, {Liang}, {Lin}, {Weng}, {Chen}, {Catinella}, {Cortese}, {Fisher}, {Ho}, {Jing}, {Jiang}, {Jiang}, {Liu}, {P{\'e}roux}, {Shao}, {Staveley-Smith}, {Wang}, and {Wang}]{Yang2025}
D.~{Yang} et al.
\newblock \emph{\apj}, 984\penalty0 (1):\penalty0 15, May 2025.
\newblock \doi{10.3847/1538-4357/adbbe8}.

\bibitem[{Yuan} et~al.(2023){Yuan}, {Krumholz}, and {Martin}]{Yuan2023}
Y.~{Yuan}, M.~R. {Krumholz}, and C.~L. {Martin}.
\newblock \emph{\mnras}, 518\penalty0 (3):\penalty0 4084--4105, Jan. 2023.
\newblock \doi{10.1093/mnras/stac3241}.

\end{thebibliography}

\end{document}